\documentclass[journal]{IEEEtran}
\usepackage[cmex10]{amsmath}
\usepackage{amssymb}
\usepackage{graphicx}
\usepackage{color}
\usepackage{comment}
\newcounter{example}
\newenvironment{example}{\vspace*{2mm} \refstepcounter{example} \emph{{Example~\theexample}:}} {\hfill$\diamond$\par\vskip1mm}
\newtheorem{theorem}{Theorem}
\newtheorem{lemma}{Lemma}
\newtheorem{corollary}{Corollary}

\makeatletter

\begin{document}
\title{A class of punctured simplex codes which are proper for error detection}
\author{Marco Baldi, Marco Bianchi, Franco Chiaraluce, Torleiv Kl{\o}ve
\thanks{Marco Baldi, Marco Bianchi, and  Franco Chiaraluce are with Dipartimento di Ingegneria dell'Informazione, Universit\`a Politecnica delle Marche, Ancona, Italy,
e-mail: \{m.baldi,m.bianchi,f.chiaraluce\}@univpm.it.
Torleiv Kl{\o}ve is with Institutt for informatikk, Universitetet i Bergen, Bergen, Norway,
e-mail: Torleiv.Klove@ii.uib.no.}}

\maketitle

\begin{abstract}
Binary linear $[n,k]$ codes that are proper for error detection are known for many combinations of $n$ and $k$.
For the remaining combinations, existence of proper codes is conjectured.
In this paper, a particular class of $[n,k]$ codes is studied in detail. In particular, it is shown that
these codes are proper for many combinations of $n$ and $k$ which were previously unsettled.
\end{abstract}


\section{Introduction}

In this paper, we study binary linear $[n,k]$ codes (codes of length $n$ and dimension $k$) used for error detection on the binary symmetric channel.
A comprehensive introduction to the field is given in \cite{K}. The basic definitions are given in Section
\ref{sub:ErrorDetection}. A main quantity is the \emph{probability of undetected error} of a code.
If the probability of undetected error is an increasing function on the interval $[0,1/2]$, the code is known
as \emph{proper} for error detection.

It is believed that proper codes exist for all lengths $n$ and dimensions $k$. However, this has been shown
only for some cases. In particular, proper $[n,k]$ codes are known to exist for any given $k$ when $n$ is sufficiently large.
The best known result in this direction was given by Kl{\o}ve and Yari \cite{KY} who showed that proper
codes exist for 
\begin{equation}
\label{KYbound}
n\ge 2^{k-1} \bigl( 2^{k-5} + 2^{\lfloor (k-5)/2\rfloor} \bigr)\mbox{ when }k\ge 5.
\end{equation}

In this paper, we study a particular class of $[n,k]$ codes where $n>2^{k-1}$. One of our results is 
that these codes are proper for many values of $n$ and $k$ where the existence of proper codes was previously unknown. 
In particular, we improve the bound (\ref{KYbound}).

We first consider $n$ in the range $2^{k-1}< n < 2^k$.  The Hamming bound proves that
 the dual of an $[n,k]$ code in this case has minimum distance at most 3. Moreover, an $[n,n-k]$ code with minimum
distance 3 can be obtained by shortening the $[2^k-1,2^k-1-k]$ Hamming code.

Two $[n,k]$ codes $C$ and $C'$ are equivalent if there exists a permutation $\pi$ of $\{1,2,\ldots ,n\}$ such that
\[C'=\{( c_{\pi(1)},c_{\pi(2)},\ldots ,c_{\pi(n)}) \mid (c_1,c_2,\ldots ,c_n)\in C\}.\]
If two codes are equivalent, then it may happen that the corresponding (repeatedly) punctured codes are not equivalent.

Let $H_k$ be some $k\times (2^k-1)$ matrix having as columns all possible nonzero vectors of length
$k$. The code generated by $H_k$ is the simplex code $S_k$, and the code having $H_k$ as parity check matrix is the well-known Hamming code.
Note that the order of the columns is not specified; all the equivalent codes are named Hamming codes.
However, when we want to puncture the code, the order is very important. 

We remind the reader that puncturing a code is equivalent to shortening the dual code.
Davydov et al. \cite{DKST} determined an ordering of the columns in $H_k$ such that any of the corresponding (repeatedly) shortened codes contains a minimal number of codewords of weight three; the shortened $[n,n-k]$ codes are obtained by removing $2^k-1-n$ columns from $H_k$ to get a $k\times n$ matrix $M_{n,k}$ and use this matrix as
the parity check matrix for the $[n,n-k]$ code. They showed that a possible choice of $M_{n,k}$ is to
have as columns the vectors that are the binary representation of the numbers from $2^k-1$ down to $2^k-n$.
For example
\begin{equation}
\label{d4}
M_{11,4}=
\begin{bmatrix}
11111111000\\
11110000111\\
11001100110\\
10101010101
\end{bmatrix}.
\end{equation}

We let $D_{n,k}$ denote the code generated by $M_{n,k}$. For our investigation, we will consider codes
that are equivalent (but not equal) to these codes; we will denote them by $S_{n,k}$. A main reason for
considering $S_{n,k}$ rather than $D_{n,k}$ is that the determination of
the weight distribution is easier for $S_{n,k}$.

In this paper, we investigate the performance of the codes $S_{n,k}$
when they are used for error detection.
We compute their weight distribution that, in turn,
permits us to calculate the undetected error probability $P_{\rm ue}(S_{n,k},p)$.
However, when the code length $n$ is large
($n\gtrsim 2^{20}$), the polynomial expressing $P_{\rm ue}(S_{n,k},p)$ may be difficult
to evaluate, even when the weight distribution is known. For this reason,
in the general case, we also find bounds on the length and dimension such that
a necessary condition for codes to be satisfactory does not hold, using
a method similar to the one proposed in \cite{XF}.

The paper is organized as follows: in Section \ref{sub:ErrorDetection}
we give some preliminaries on error detection; in Section \ref{sub:DKSTConstruction}
we describe the construction of $S_{n,k}$ and show that $S_{n,k}$ and $D_{n,k}$ are equivalent; in Section \ref{sec:Weights-Distribution}
we determine the weight distribution of $S_{n,k}$; in Section \ref{sec:Prob} we study the undetected error probability
of $S_{n,k}$ and its dual code; in Section \ref{sec:Approximate-Analysis}
we give an asymptotic analysis; in Section \ref{sec:Generalization} we give a generalization of the construction
to lenghts $n\ge 2^{k}$, finally, in Section \ref{sec:Results}
we summarize our results.


\section{Error detection
\label{sub:ErrorDetection}}

We start by defining $P_{\rm ue}(C,p)$, the undetected error probability
for an $[n,k]$ code $C$ when used on the binary symmetric channel with error probability $p$:
\begin{equation}
P_{\rm ue}(C,p)=\sum_{w=1}^{n}A_{w}p^{w}(1-p)^{n-w},
\label{eq:PueCode}
\end{equation}
where $A_{w}$ is the number of codewords having Hamming weight $w$, see e.g. \cite[Section 2.1.2]{K}. 

One can also express this polynomial in terms of the weight distribution of the dual code, see e.g.
\cite[Theorem 2.4]{K}. If $A^{\bot}_{w}$
is the number of codewords having Hamming weight $w$ in the dual code $C^{\bot}$, we have:
\begin{equation}
P_{\rm ue}(C,p)=2^{k-n}\sum_{w=0}^{n}A^{\bot}_{w}(1-2p)^{w}-(1-p)^{n}.
\label{eq:PueDual}
\end{equation}

As mentioned in the introduction, if $P_{\rm ue}(C,p)$ is an increasing function on $\bigl[0,\frac{1}{2}\bigr]$,
the code $C$ is called \emph{proper} for error detection. If
\[P_{\rm ue}(C,p)\leq P_{\rm ue}(C,1/2)\]
for every $p \in \bigl[0,\frac{1}{2}\bigr]$, $C$ is
called \emph{good} for error detection. If
\[P_{\rm ue}(C,p)\leq 2^{k-n}\]
for every $p \in \bigl[0,\frac{1}{2}\bigr]$, $C$ is
called \emph{satisfactory} for error detection, see \cite[p. 38]{K}.
A code that is not satisfactory is called \emph{ugly}.
When a code is proper
then it is satisfactory; so, if it is ugly it
is clearly not proper (nor good).


\section{The code construction
\label{sub:DKSTConstruction}}

We first describe a particular parity check matrix $H_k$ for the Hamming code.
  For $0\le m \le k-1$, let $H_k^{(m)}$ be the $k\times 2^m$ matrix constructed as follows:
\begin{itemize}
\item The first $k-m-1$ rows are all-zero vectors.
\item Row $k-m$ is the all-one vector.
\item In the $m\times 2^m$ matrix consisting of the last $m$ rows,
the columns are ordered lexicographically.
\end{itemize}
Then
\[H_{k}=\Bigl[H_{k}^{(k-1)}\,|\, H_{k}^{(k-2)}\,|\,\ldots \,|\, H_{k}^{(0)}\Bigr].\]

We illustrate this with an example. For $k=4$, we get
\[
H_{4}^{(0)}= \begin{bmatrix}
0\\
0\\
0\\
1
\end{bmatrix},
H_{4}^{(1)}= \begin{bmatrix}
00 \\
00 \\
11 \\
01
\end{bmatrix},
H_{4}^{(2)}= \begin{bmatrix}
0000 \\
1111 \\
0011 \\
0101
\end{bmatrix},\]
\[H_{4}^{(3)}= \begin{bmatrix}
11111111 \\
00001111 \\
00110011 \\
01010101
\end{bmatrix}
\]
and so
\begin{align*}
H_{4}=& \Bigl[H_{4}^{(3)}|H_{4}^{(2)}|H_{4}^{(1)}|H_{4}^{(0)}\Bigr] \\
=&
\begin{bmatrix}
111111110000000\\
000011111111000\\
001100110011110\\
010101010101011
\end{bmatrix}.
\end{align*}

We let $H_k(n)$ denote the $k\times n$ matrix containing the first $n$ columns of $H_k$.
For example
\begin{equation}
\label{h4}
H_4(11)=\begin{bmatrix}
11111111000\\
00001111111\\
00110011001\\
01010101010
\end{bmatrix}.
\end{equation}

We let $S_{n,k}$ denote the code generated by $H_k(n)$.
We see that $S_{2^{k-1},k}$ is the first order Reed-Muller code
and $S_{2^k-1,k}=S_k$, the simplex code. Both of these codes are known to be proper
(and this is easy to show).
The Hamming code is $S_{2^k-1,k}^\perp$.
The code having $H_k(n)$ as parity check matrix is a shortened Hamming code which we denote by $C_{n,n-k}$. We note that
$C_{n,n-k}=S_{n,k}^\perp$.
In the rest of the paper (except Section \ref{sec:Generalization}) we will assume that
$2^{k-1}< n \le 2^k-1$.

\begin{theorem}

\label{DKST-BS}
The codes $S_{n,k}$ and $D_{n,k}$ are equivalent.
\end{theorem}
\begin{IEEEproof}
We first illustrate by the example $k=4$ and $n=11$, that is, the matrices (\ref{h4}) and (\ref{d4}).
Adding the second row in (\ref{h4}) to the third and forth rows, we get
\begin{equation}
\label{hh4}
\begin{bmatrix}
11111111000\\
00001111111\\
00111100110\\
01011010101
\end{bmatrix}.
\end{equation}
This is an alternative generator matrix for $S_{11,4}$.
The last three columns are the same in (\ref{hh4}) and (\ref{d4}), and the first eight columns of (\ref{hh4}) are a permutation of the first eight columns in (\ref{d4}). Hence, $S_{11,4}$ and $D_{11,4}$ are equivalent.

In the general case, if $n\in[2^k-2^{k-m}+1,2^k-2^{k-m-1}-1]$ for some $m$, $1\le m \le k-1$, we add row $m+1$ in $H_k(n)$ to all the rows below. This gives an alternative generator
matrix $H'_k(n)$ for $S_{n,k}$. 
The first $2^{k-1}$ columns of $H'_k(n)$ are a permutation of the binary representations of 
$i\in [2^{k-1},2^k-1]$,
the next $2^{k-2}$ columns of $H'_k(n)$ are a permutation of the binary representations of $i\in [2^{k-2},2^{k-1}-1]$, etc.
The final $n-2^k+2^{k-m}$ columns in $H'_k(n)$ and $M_{n,k}$ are the same.
Hence, $S_{n,k}$ and $D_{n,k}$ are equivalent.

If $n=2^k-2^{k-m}$ for some $m$, $1\le m \le k$, the same argument shows that the columns of 
$H_k(n)$ are a permutation of the columns of $M_{n,k}$, and so again $S_{n,k}$ and $D_{n,k}$ are equivalent.
\end{IEEEproof}


\section{Weight Distribution of $S_{n,k}$
\label{sec:Weights-Distribution}}

The main question we consider is: for which $n$ and $k$ is $S_{n,k}$
proper for error detection? We will also in some cases consider the simpler question: 
for which $n$ and $k$ is $S_{n,k}$ satisfactory for error detection? 

We note that this is equivalent to the question: 
for which $n$ and $k$ is $C_{n,n-k}$ satisfactory for error detection?
The reason is the following known lemma.
\begin{lemma}\cite[Theorem 2.8]{K}.
\label{sdual}
A code is satisfactory if and only if the dual code is satisfactory.
\end{lemma}

To determine the probability of undetected error for $S_{n,k}$, we have to determine its weight
distribution. This is done in this section. We break the argument down into a number of lemmas.

We first give some further notations. We observe that the matrix
\[\Bigl[H_{k}^{(k-1)}\,|\, H_{k}^{(k-2)}\,|\,\ldots \,|\, H_{k}^{(k-m)}\Bigr]\]
has length
\[ \sum_{j=1}^{m}2^{k-j} = 2^k-2^{k-m} .\]

  For a given $n$, let $m$ be determined by
\[ 2^k-2^{k-m} <n\le2^k-2^{k-m-1}. \]
Since $2^{k-1}<n<2^k$, we have $1\le m \le k-1$.
Let $(\alpha_1, \alpha_2, \ldots , \alpha_k)$ denote the last column of $H_k(n)$.

\begin{lemma}
\label{alpha-lemma}
Let $2^k-2^{k-m} <n\le2^k-2^{k-m-1}$. Then
\begin{equation}
\label{alm} \alpha_1=\ldots=\alpha_m=0, \quad\alpha_{m+1}=1,
\end{equation}
and $\alpha_{m+2},\ldots,\alpha_k$ are determined by
\begin{equation}
\label{alr}
\sum_{i=0}^{k-m-2} \alpha_{k-i}\, 2^i = n-1- 2^k+2^{k-m}.
\end{equation}
\end{lemma}

\begin{IEEEproof}
The last column in $H_k(n)$ is a column in $H_k^{(k-m-1)}$. Hence (\ref{alm}) follows. Moreover,
its number in $H_k^{(k-m-1)}$ is $n-1- (2^k-2^{k-m})$
when we count the first column as number zero.
The columns in $H_k^{(k-m-1)}$ are ordered lexicographically and so (\ref{alr}) follows.
\end{IEEEproof}

Let $w_i$ denote the weight of the $i$-th row in $H_k(n)$. As usual,
$\lfloor x \rfloor$ denotes the largest integer less than or equal to $x$.

\begin{lemma}
\label{w-lemma1}
Let $2^k-2^{k-m} <n\le2^k-2^{k-m-1}$. Then
\begin{equation}
\label{wm}
w_1 = \cdots = w_m = 2^{k-1}
\end{equation}
and
\begin{equation}
\label{wm1}
w_{m+1}=n- 2^{k-1}+2^{k-m-1}.
\end{equation}
If $m+2\le i \le k$ and $\alpha_i=0$, then
\begin{equation}
\label{wm2}
w_i= 2^{k-i} \Bigl\lfloor\frac{n-1}{2^{k-i+1}} \Bigr\rfloor.
\end{equation}
If $m+2\le i \le k$ and $\alpha_i=1$, then
\begin{equation}
\label{wm3}
w_i= n - 2^{k-i} \Bigl\lfloor \frac{n-1}{2^{k-i+1}} \Bigr\rfloor - 2^{k-i}.
\end{equation}
\end{lemma}

\begin{IEEEproof}
All the rows of $H_k$ have weight $2^{k-1}$. The first $m$ rows of $H_k(n)$ are obtained
from rows in $H_k$ removing some zeros.
Hence $w_i=2^{k-1}$ for $1\le i \le m$.

Before we go on with the proof, let us take a closer look at $H_k^{(m)}$. Row $i>m$ consists of
consecutive blocks of zeros and ones, each block of length $2^{k-i}$.
We use the term \emph{double block} for a zero-block combined with the following one-block;
it has length $2^{k-i+1}$. Now, let
\begin{equation}
\label{nu}
n = \Bigl\lfloor \frac{n-1}{2^{k-i+1}} \Bigr\rfloor 2^{k-i+1} + \nu \mbox{ where }
1\le \nu \le 2^{k-i+1} .
\end{equation}
Then row $i$ in $H_k(n)$ consists of
$\Bigl\lfloor\frac{n-1}{2^{k-i+1}} \Bigr\rfloor$
double blocks of length $2^{k-i+1}$, each of weight
$2^{k-i}$, followed by an incomplete double block of length $\nu$ that has to be considered further
(when $\nu=2^{k-i+1}$, the incomplete double block is, of course, a full double block).

If $\alpha_i=0$, the incomplete double block is all zero, and so (\ref{wm2}) follows.

If $\alpha_i=1$, the incomplete double block consists of a full block (of length $2^{k-i}$) of zeros
followed by an incomplete block of ones of length
$\nu-2^{k-i}$. Hence
\begin{align*}
w_i &= 2^{k-i} \Bigl\lfloor\frac{n-1}{2^{k-i+1}} \Bigr\rfloor + \nu-2^{k-i} \\
&= 2^{k-i} \Bigl\lfloor\frac{n-1}{2^{k-i+1}} \Bigr\rfloor + n - \Bigl\lfloor
\frac{n-1}{2^{k-i+1}} \Bigr\rfloor 2^{k-i+1} -2^{k-i}\\
&= n - 2^{k-i} \Bigl\lfloor\frac{n-1}{2^{k-i+1}} \Bigr\rfloor -2^{k-i}.
\end{align*}
This proves (\ref{wm3}). The proof of (\ref{wm1}) is similar (and even simpler).
\end{IEEEproof}

\begin{lemma}
\label{lem1}
Consider sums of rows from $H_k(n)$.

a) Any of the $2^m-1$ non-zero sums of some of the first $m$ rows have weight $2^{k-1}$.

b) Any of the $2^{m}$ sums containing row $m+1$ and zero or more previous rows have weight $w_{m+1}$.

c) For $m+2\le i\le k$, $2^{i-2}$ sums containing row $i$ and some previous rows have weight
$w_{i}$ and the other $2^{i-2}$ sums have weight $n-w_{i}$.
\end{lemma}

\begin{IEEEproof}
  For each sum of rows from the first $m$, the corresponding sum of rows in $H_k$
are codewords in the simplex code $S_k$. These always have weight $2^{k-1}$. Since only zeros have been
removed to get the corresponding rows in $H_k(n)$, their sum also has weight $2^{k-1}$.
This proves a).

Let $i\ge m+2$. We note that in the set of positions of a double block in row $i\ge m+2$
in $H_k^{(k-j)}$ for $j\le m+1$, the elements of
any previous row are all zero or all one. Therefore, the weight of these positions
in any sum of row $i$ and a combination of previous rows is $2^{k-i}$.

It remains to consider the contribution to the weight from the last $\nu$ positions
(where $\nu$ is defined by (\ref{nu})).

Case I) $\alpha_i=0$. In this case, all the last $\nu$ elements of row $i$ are zeros.
Any previous row has all zeros or all ones in these positions,
and so the weight of the elements in these positions in any sum is either 0 or $\nu$.
Hence, the weight of the sum is either $w_i$ or $n-w_i$, where $w_i$ is given by (\ref{wm2}).
Moreover, row $m+1$ has all
ones in the last $\nu$ positions. Hence, half of the
$2^{i-1}$ sums has weight $w_i$ and
the other half has weight $n-w_i$.

Case II) $\alpha_i=1$. In this case, all the last $\nu$ elements of row $i$ are
$2^{k-i}$ zeros followed by $\nu-2^{k-i}$ ones. The weight of the last $\nu$
elements in a sum is therefore $\nu-2^{k-i}$ or $2^{k-i}$. Hence, the weight of
a sum is $w_i$ or $n-w_i$, where now $w_i$ is given by (\ref{wm3}).
As done above, considering sums containing row $m+1$, we can see that
the multiplicities of these two weights are the same.
This proves c).

  Finally, consider row $m+1$. Any previous row has all zeros in the last $\nu$ positions.
Hence, the weight of any sum involving row $m+1$ and previous rows is $w_{m+1}$.
This proves b).
\end{IEEEproof}

We next give an alternative expression for $w_i$.

\begin{lemma}
\label{w-alt}
Let $2^k-2^{k-m} <n\le2^k-2^{k-m-1}$. \\
a) If $m+2\le i \le k$ and $\alpha_i=0$, then
\[w_{i}= 2^{k-1}-2^{k-m-1}+\sum_{j=m+2}^{i-1} \alpha_j \,2^{k-1-j} .\]
b) If $m+2\le i \le k$ and $\alpha_i=1$, then
\[w_{i}= 2^{k-1}-2^{k-m-1}+ 1+ \sum_{j=m+2}^{i-1} \alpha_j \,2^{k-1-j}+ \sum_{j=i+1}^{k} \alpha_j \,2^{k-j} .\]
c) Further,
\[w_{m+1}=2^{k-1}-2^{k-m-1}+ 1+ \sum_{j=m+2}^{k} \alpha_j \,2^{k-j}.\]
\end{lemma}

\begin{IEEEproof}
 From (\ref{alr}) we get
\[n-1 = 2^k - 2^{k-m} + \sum_{j=m+2}^{k} \alpha_{j}\, 2^{k-j} .\]
Hence
\[ \frac{n-1}{2^{k-i+1}} = 2^{i-1}- 2^{i-m-1} + \sum_{j=m+2}^{i-1} \alpha_{j}\, 2^{i-1-j} +r\]
where
\[r= \sum_{j=i}^{k} \alpha_{j}\, 2^{i-1-j} \le \sum_{j=i}^{k} 2^{i-1-j} = 1- 2^{i-1-k}<1 .\]
Hence
\[ \Bigl\lfloor \frac{n-1}{2^{k-i+1}} \Bigr\rfloor
= 2^{i-1}- 2^{i-m-1}
  + \sum_{j=m+2}^{i-1} \alpha_{j}\, 2^{i-1-j} \]
and so, by (\ref{wm2}),
\[w_i= 2^{k-i} \Bigl\lfloor
  \frac{n-1}{2^{k-i+1}}\Bigr\rfloor = 2^{k-1}- 2^{k-m-1}
  + \sum_{j=m+2}^{i-1} \alpha_{j}\, 2^{k-1-j} .\]
This proves a).

Similarly, if $\alpha_{i}=1$, (\ref{wm3}) gives
\begin{align*}
w_i &= n- 2^{k-i} \Bigl\lfloor \frac{n-1}{2^{k-i+1}}\Bigr\rfloor - 2^{k-i} \\
&= 1+ 2^k - 2^{k-m} + \sum_{j=m+2}^{k} \alpha_{j}\, 2^{k-j} - 2^{k-i}\\
& \quad - 2^{k-1}+ 2^{k-m-1} - \sum_{j=m+2}^{i-1} \alpha_{j}\, 2^{k-1-j} \\
&= 1+ 2^{k-1}- 2^{k-m-1} + \sum_{j=m+2}^{i-1} \alpha_{j}\, 2^{k-1-j} \\
&\quad + \alpha_i\, 2^{k-i} + \sum_{j=i+1}^{k} \alpha_{j}\, 2^{k-j} - 2^{k-i} \\
&= 1+ 2^{k-1}- 2^{k-m-1} + \sum_{j=m+2}^{i-1} \alpha_{j}\, 2^{k-1-j}\\
&\quad + \sum_{j=i+1}^{k} \alpha_{j}\, 2^{k-j},
\end{align*}
since $\alpha_i=1$.
This proves b).
  Finally, c) follows directly by substituting the expression for $n$ in the expression for
$w_{m+1}$ in (\ref{wm1}).
\end{IEEEproof}

\begin{example}
\label{ex0}
Consider $n=2^k-2^{k-m-1}$, where $1\le m < k$. We have
\[\alpha_i=0 \mbox{ for }1\le i \le m \mbox{ and }
  \alpha_i=1 \mbox{ for }m+1\le i \le k.\]
Using Lemma \ref{w-alt}, we see that $S_{n,k}$ has $2^m-1$ codewords of weight $w_1=2^{k-1}$ and $2^k-2^m$
codewords of weight
\[2^{k-1}-2^{k-m-2}=n/2. \]
In particular, the minimum distance is $n/2$.
Hence, the code $S_{n,k}$ is proper
 (see \cite[Theorem 2.2]{K}).
\end{example}

\begin{lemma}
\label{w-diff}
Let \[2^k-2^{k-m} <n<2^k-2^{k-m-1}\]
 and $m+2\le i \le k$.

a) If $\alpha_i=\alpha_{i+1}$, then $w_{i+1}=w_{i}$.

b) If $\alpha_i=0$ and $\alpha_{i+1}=1$, then $w_{i+1}>w_{i}$.

c1) If $\alpha_i=1$, $\alpha_{i+1}=0$, and $\alpha_j=1$ for all $j$ such that $i+2\le j \le k$,
then $w_{i+1}=w_{i}$.

c2) If $\alpha_i=1$, $\alpha_{i+1}=0$, and $\alpha_j=0$ for at least one $j\ge i+2$, then $w_{i+1}>w_{i}$.

d) In all cases,
\[w_1\ge w_{m+1}>w_k\ge w_{k-1}\ge w_{k-2}\ge \cdots \ge w_{m+2}.\]
In particular, the minimum distance $d$ of $S_{n,k}$ is $w_{m+2}$.

e) $w_{m+1}> n/2$.
\end{lemma}

\begin{IEEEproof}
a) If $\alpha_i=\alpha_{i+1}=0$, then Lemma \ref{w-alt} gives
\[w_{i+1}-w_i = \alpha_{i}\, 2^{k-1-i} =0.\]
If $\alpha_i=\alpha_{i+1}=1$, then Lemma \ref{w-alt} gives
\[w_{i+1}-w_i = \alpha_i\, 2^{k-i-1} - \alpha_{i+1}\, 2^{k-(i+1)} =0.\]

b) If $\alpha_i=0$ and $\alpha_{i+1}=1$, then Lemma \ref{w-alt} gives
\[w_{i+1}-w_i = 1+ \sum_{j=i+2}^{k} \alpha_j \,2^{k-j}>0 .\]

c) If $\alpha_i=1$ and $\alpha_{i+1}=0$, then Lemma \ref{w-alt} gives
\[w_{i+1}-w_i = 2^{k-1-i}-1- \sum_{j=i+2}^k \alpha_j\, 2^{k-j} .\]
We have
\[\sum_{j=i+2}^k \alpha_j\, 2^{k-j} \le \sum_{j=i+2}^k 2^{k-j}=2^{k-1-i}-1\]
with equality if and only if $\alpha_j=1$ for
$i+2\le j \le k$.

d) We have
\[\sum_{j=m+2}^{k} \alpha_j \,2^{k-j}\le
\sum_{j=m+2}^{k} 2^{k-j} = 2^{k-m-1}-1, \]
and so
\[w_{m+1} \le 2^{k-1}- 2^{k-m-1}+ 1+ 2^{k-m-1}-1=w_1.\]
 Further, both for $\alpha_k=0$ and $\alpha_k=1$, Lemma \ref{w-alt} gives
\[w_{m+1}-w_k = 1+ \sum_{j=m+2}^{k-1} \alpha_j\, 2^{k-j-1} >0.\]
 For $m+2\le i \le k$, a), b), c1), and c2) show that $w_i\ge w_{i-1}$.

e) Equation (\ref{wm1}) implies that
\[n-2w_{m+1}=2^k-2^{k-m}-n<0.\]
\end{IEEEproof}

Let
\begin{align}
n(k,m) &=2^k-2^{k-m}+2^{k-m-2} =2^k-2^{k-m-1}-2^{k-m-2} \nonumber \\
&=2^k-3\cdot 2^{k-m-2}=2^{k-m-2}(2^{m+2}-3), \label{nrm}
\end{align}
the midpoint of the interval
$[2^k-2^{k-m},2^k-2^{k-m-1}]$.

\begin{lemma}
\label{dmin}
Let $d$ be the minimum distance of $S_{n,k}$.

a) If $2^{k}-2^{k-m} \le n\le n(k,m)$, then
\[d= 2^{k-1}-2^{k-m-1}.\]

b) If $n(k,m) \le n\le2^{k} -2^{k-m-1}$, then
\[n-d= 2^{k-1} -2^{k-m-2}.\]
\end{lemma}

\begin{IEEEproof}
a) We have
\[n-1-2^k+2^{k-m}\le 2^{k-m-2}-1.\]
Hence $\alpha_{m+2}=0$. By Lemma \ref{w-alt}a),
\[ w_{m+2}=2^{k-1}-2^{k-m-1}.\]

b) We have $ \alpha_{m+2} =1$.  From (\ref{alr}) and Lemma \ref{w-alt}b),
\begin{align*}
n-w_{m+2}& =2^{k-1}-2^{k-m-1}+\alpha_{m+2}\, 2^{k-m-2}\\
& =2^{k-1}-2^{k-m-2}.
\end{align*}
\end{IEEEproof}

\begin{lemma}
\label{Admin}
a) If $w_{m+2}= w_{k}$, then
\[A_d=2^{k-1}-2^{m}.\]

b) If $w_{m+2}< w_{k}$ and $i\ge m+2$ is given by 
\[w_{m+2}= w_{i} < w_{i+1},\]
 then
\[A_d=2^{i-1}-2^{m}.\]
In particular, $A_d\ge 2^m$ in all cases.
\end{lemma}

\begin{IEEEproof}
The conditions imply that $w_j$ has value $d$ exactly for $m+2\le j \le i$ (where $i=k$ for
case a)). Hence
\[A_d=\sum_{j=m+2}^i 2^{j-2}= 2^{i-1}-2^m.\]
\end{IEEEproof}


\section{Probability of undetected error of $S_{n,k}$ and $C_{n,n-k}$
\label{sec:Prob}}

 From (\ref{eq:PueCode}), (\ref{eq:PueDual}), and Lemma \ref{lem1}, we get the following theorems.

\begin{theorem}
\label{th1}
Let $2^k-2^{k-m} <n\le2^k-2^{k-m-1}$. Then
\begin{eqnarray*}
\lefteqn{P_{\rm ue}(S_{n,k},p)= (2^m-1) p^{w_1}(1-p)^{n-w_1}} \\
&& +\, 2^{m} p^{w_{m+1}} (1-p)^{n-w_{m+1}}\\
&& + \sum_{i=m+2}^k 2^{i-2} \Bigl\{ p^{w_i} (1-p)^{n-w_i} + p^{n-w_i} (1-p)^{w_i}\Bigr\}.
\end{eqnarray*}
\end{theorem}

\begin{theorem}
\label{th1dual}
Let $2^k-2^{k-m} <n\le2^k-2^{k-m-1}$. Then
\begin{eqnarray*}
\lefteqn{ 2^k P_{\rm ue}(C_{n,n-k},p)= (2^m-1) (1-2p)^{w_1} }\\
&& +\, 2^{m} (1-2p)^{w_{m+1}}\\
&& + \sum_{i=m+2}^k 2^{i-2} \Bigl\{ (1-2p)^{w_i} + (1-2p)^{n-w_i}\Bigr\}.
\end{eqnarray*}
\end{theorem}

\begin{example}
\label{ex1}
Consider $n=n(k,m)= 2^{k}-3\cdot 2^{k-m-2}$ where $k\ge m+3$. Then
\begin{align*}
\alpha_i&=0, \mbox{ for } 1\le i \le m,\\
\alpha_{m+1} &= 1, \\
\alpha_{m+2} &= 0, \\
\alpha_i&= 1 \mbox{ for } m+3\le i \le k.
\end{align*}
Using Lemmas \ref{w-lemma1} and \ref{w-alt} we get
\begin{align*}
w_i&= 2^{k-1} \mbox{ for } 1\le i \le m, \\
w_{m+1}&= 2^{k-1}-2^{k-m-2}, \\
w_{m+2}&= 2^{k-1}-2^{k-m-1}, \\
n-w_{m+2}&= 2^{k-1}-2^{k-m-2}, \\
n-w_i &= w_i= 2^{k-1}-3\cdot 2^{k-m-3} \mbox{ for } m+3\le i \le k.
\end{align*}
Hence, the weight distribution of $S_{n,k}$ is given by Table \ref{tabnr}.
\begin{table}[!t]
\caption{ The weight distribution of $S_{2^{k}-3\cdot 2^{k-m-2},k}$ }
\label{tabnr}
\[\begin{array}{|l|l|} \hline
w & A_w \\ \hline
0& 1 \\
2^{k-1}-2^{k-m-1} & 2^m \\
2^{k-1}-3\cdot 2^{k-m-3} & 2^k-2^{m+2} \\
2^{k-1}-2^{k-m-2} & 2^{m+1} \\
2^{k-1}& 2^m-1\\ \hline
\end{array}\]\end{table}
\end{example}

\begin{example}
  For $k=9$ and $m=1$ in Example \ref{ex1} we get $n=320$ and
\begin{align*}
P_{\rm ue}(S_{320,9},p)=&\, 2\, p^{128}(1-p)^{192} + 504\, p^{160}(1-p)^{160} \\
&\, + 4\, p^{192}(1-p)^{128} + p^{256}(1-p)^{64}.
\end{align*}
In  Fig. \ref{s320} we give the graphs of $P_{\rm ue}(S_{320,9},p)$ and the terms
$2\, p^{128}(1-p)^{192}$ and $504\, p^{160}(1-p)^{160}$. The contributions from the last two terms,
$4\, p^{192}(1-p)^{128}$ and $p^{256}(1-p)^{64}$
are so small that they are not visible on the graph.
The graph illustrates that $S_{320,9}$ is ugly.
  For small $p$ ($p$ up to approximately 0.42), $2\, p^{128}(1-p)^{192}$
is the dominating term; in this region the difference
\[P_{\rm ue}(S_{320,9},p)-2\, p^{128}(1-p)^{192}\]
is so small that it is not visible on the graph.
  For $p$ close to 0.5, the term $504\, p^{160}(1-p)^{160}$ dominates.
\begin{figure}[!t]
\begin{center}
\includegraphics[scale=0.45]{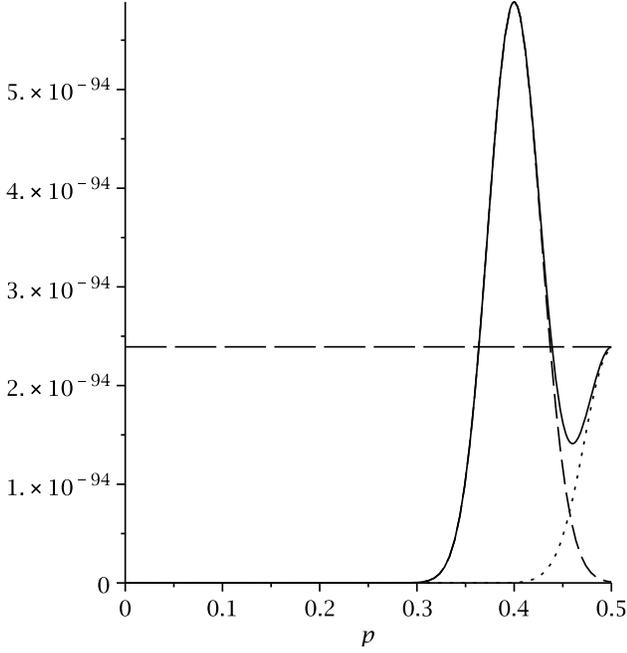}
\caption{Plot of $P_{\rm ue}(S_{320,9},p)$ (solid line), $2\, p^{128}(1-p)^{192}$ (dashed line),
$504\, p^{160}(1-p)^{160}$ (dotted line), and $2^{9-320}$ (long dashed).}
\label{s320}
\end{center}
\end{figure}
\end{example}

Theorem \ref{th1} can be used to determine if the code $S_{n,k}$ is proper
and Theorem \ref{th1dual} if the code $C_{n,n-k}$ is proper.
We just compute $\frac{dP_{\rm ue}}{dp}$ and check
the presence or absence of real roots in $(0,\frac{1}{2})$.
  For moderate values of $n$ and $k$ (e.g. $k \lesssim 20$), this is feasible in a reasonable time.

Before we give a main general result, we quote two lemmas from \cite{K}.

\begin{lemma}
\label{incr}
a) \cite{K} Theorem 2.2: if $w\ge n/2$, then
\[p^w(1-p)^{n-w}\]
is increasing on $[0,1/2]$.

b) \cite{K} Lemma 3.5: if $(n-\sqrt{n})/2\le w \le n/2$, then
\[p^w(1-p)^{n-w}+p^{n-w}(1-p)^w\]
is increasing on $[0,1/2]$.
\end{lemma}

Let
\[ \overline{\tau}_{k,m} =\min \Bigl\{2^{k-m-2},\frac{1+\sqrt{1+2^{k+2}-2^{k-m+2}} }{2}\Bigr\},\]
\[\underline{\tau}_{k,m}=\min \Bigl\{2^{k-m-2}-1,\frac{-1+\sqrt{1+2^{k+2}-2^{k-m+1}} }{2}\Bigr\}.\]

\begin{theorem}
\label{S-proper}
  For $k>m\ge 1$, if
\begin{equation}
\label{kcon1}
2^k-2^{k-m} \le n \le 2^k-2^{k-m}+ \overline{\tau}_{k,m}
\end{equation}
or
\begin{equation}
\label{kcon2}
2^k-2^{k-m-1}-\underline{\tau}_{k,m} \le n \le 2^k-2^{k-m-1},
\end{equation}
then $S_{n,k}$ is proper.
\end{theorem}

\begin{IEEEproof}
  For $n=2^k-2^{k-m}$ and $n=2^k-2^{k-m-1}$, $S_{n,k}$ is proper by Example \ref{ex0}.
  For $2^k-2^{k-m}<n<2^k-2^{k-m-1}$, (\ref{wm}), Lemma \ref{w-diff}d),
and Lemma \ref{w-diff}e) imply that
\[w_1=\cdots =w_m\ge w_{m+1}>n/2\]
and so
$p^{w_i}(1-p)^{n-w_i}$ is increasing on $[0,1/2]$ for $1\le i \le m+1$ by Lemma \ref{incr}a).

Now, consider $n= 2^k-2^{k-m} +\eta $, where $0<\eta\le 2^{k-m-2}$.
By Lemma \ref{dmin}a)
\[d= 2^{k-1}-2^{k-m-1}<\frac{n}{2}.\]
By Lemmas \ref{w-diff}d) and \ref{incr}b),
\[ p^{w_i}(1-p)^{n-w_i} + p^{n-w_i}(1-p)^{w_i} \]
is increasing for $m+2\le i \le k$ if $2d\ge n-\sqrt{n}$, that is, if
\[2^{k}-2^{k-m}=2d \ge2^k-2^{k-m} +\eta - \sqrt{2^k-2^{k-m} +\eta } .\]
This is equivalent to
\[\sqrt{ 2^k-2^{k-m} +\eta } \ge \eta\]
and
\[2^k-2^{k-m} +\eta \ge \eta^2.\]
Solving this for $\eta$, we get
\[\eta \le \frac{1+\sqrt{1+4(2^k-2^{k-m})}}{2}.\]
We see that if (\ref{kcon1}) is satisfied, then all the terms in
$P_{\rm ue} (S_{n,k},p)$ are increasing on $[0,1/2]$. Consequently, $S_{n,k}$ is proper.

Next, let
$ n= 2^k-2^{k-m-1} -\eta$ where $0< \eta \le 2^{k-m-2}-1$. Then, by Lemma \ref{dmin}b),
\[d= 2^{k-1}-2^{k-m-2} -\eta.\]
We want
\begin{align*}
2^k-2^{k-m-1} -2\eta &=2d \\
&\ge 2^k-2^{k-m-1} -\eta -\sqrt{2^k-2^{k-m-1} -\eta }.
\end{align*}
Solving for $\eta$, we get $\eta^2+\eta\le 2^2-2^{k-m-1}$ and so
\[\eta \le \frac{-1+\sqrt{1+4(2^k-2^{k-m-1})}}{2}. \]
As above, if (\ref{kcon2}) is satisfied, then $S_{n,k}$ is proper.
\end{IEEEproof}

When $S_{n,k}$ is proper, then it is satisfactory, and so, by Lemma \ref{sdual}, $C_{n,n-k}$ is satisfactory.
Hence we get the following corollary.

\begin{corollary}
\label{c-kange}
If $n$ is in the range defined by (\ref{kcon1}) or (\ref{kcon2}) for some $m\ge 1$, then
$C_{n,n-k}$ is satisfactory.
\end{corollary}

\begin{theorem}
\label{p-kange}
a) If
\begin{equation}
\label{mb}
m\ge \Bigl\lceil \frac{k-3}{2}\Bigr\rceil,
\end{equation}
then $S_{n,k}$ is proper for all
$n\in \Bigl[2^k-2^{k-m},2^k-2^{k-m-1}\Bigr]$.

b) $S_{n,k}$ is proper for all $n\in \Bigl[2^k-2^{k-\bigl\lceil \frac{k-3}{2}\bigr\rceil},2^k-1\Bigr]$.
\end{theorem}

\begin{IEEEproof}
We have $\overline{\tau}_{k,m}= 2^{k-m-2} $ if and only if
\begin{equation}
\label{full}
\frac{1+\sqrt{1+2^{k+2}-2^{k-m+2}} }{2} \ge 2^{k-m-2}.
\end{equation}
We observe that $2^{k-m-2}$ decreases with increasing $m$ and $\sqrt{1+2^{k+2}-2^{k-m+2}}$
increases with increasing $m$.

Let $x=2^{k-m-1}$. Then (\ref{full}) is equivalent to the following sequence of inequalities
\[1+\sqrt{1+2^{k+2}-8x} \ge x,\]
\[1+2^{k+2}-8\,x\ge (x-1)^2= x^2-2\,x+1,\]
\begin{equation}
\label{full2}
x^2+6\, x \le 2^{k+2}.
\end{equation}
  For $k=2\,m+3$ we get $x=2^{m+2}$ and so
\[x^2+6\,x = 2^{2\,m+4}+6\cdot 2^{m+2}\le 2^{2\,m+5}=2^{k+2}\]
for all $m\ge 1$.
However, for $k=2\,m+4$ we get $x=2^{m+3}$ and so
\[x^2+6\,x = 2^{2\,m+6}+6\cdot 2^{m+3}>2^{2\,m+6}=2^{k+2}.\]
Hence, (\ref{full}) is satisfied if and only if $k\le 2\,m+3$, that is when (\ref{mb}) is satisfied.
Therefore, if (\ref{mb}) is satisfied, then $S_{n,k}$ is proper for all
$n\in [2^k-2^{k-m},2^k-2^{k-m}+2^{k-m-2}]$.

Next, since $-2^{k-m+1}>-2^{k-m+2}$, we see that if (\ref{full}) is satisfied, then
\begin{align*}
\frac{-1+\sqrt{1+2^{k+2}-2^{k-m+1}} }{2} &> \frac{1+\sqrt{1+2^{k+2}-2^{k-m+2}} }{2}-1 \\
&\ge 2^{k-m-2}-1.
\end{align*}
Hence, $\underline{\tau}_{k,m}=2^{k-m-2}-1$, and so $S_{n,k}$ is proper also for all
\[n\in [2^k-2^{k-m}+2^{k-m-2}+1,2^k-2^{k-m-1}].\]
This, combined with the result above, proves a).

Since $S_{n,k}$ is proper for
\[n\in [2^k-2^{k-m},2^k-2^{k-m-1}]\]
for all $m\ge \Bigl\lceil \frac{k-3}{2}\Bigr\rceil$, b) follows.
\end{IEEEproof}

  Based on the previous theorems, we have found a set of values of $n$ for which $S_{n,k}$ is proper and, hence,
satisfactory. For other values of $n$, the existence of real roots of
$\frac{dP_\mathrm{ue}}{dp}$ in $\left(0,\frac{1}{2}\right)$ must
be checked. However, for large values of $k$  and $n$ in general it may be difficult to numerically compute
the polynomial's real roots, or even just to determine the existence of real roots
(e.g. using Sturm's chain). However, in many cases we can decide that the code $S_{n,k}$ (and hence
$C_{n,n-k}$) is not satisfactory (i.e. ugly) by showing that $P_{\rm ue}(C,p) >2^{k-n}$
for some value of $p$.
How should the value of $p$ be chosen? There is no theory that can give an exact answer to this question.
However, it is known that if the minimum distance of the code is $d$, then $A_d\, p^d (1-p)^{n-d}$
is often the dominating
term of $P_{\rm ue}(C,p)$, except for large $p$. This is well illustrated by the example of $S_{320,9}$ given
in  Fig. \ref{s320}. Since $p^d (1-p)^{n-d}$ has
its maximum for $p=d/n$, a good choice for $p$ may be $p=d/n$. This gives the following
sufficient condition for $S_{n,k}$ to be ugly:

\begin{equation}
\label{adr}
2^{k-n}< A_d\, \Bigl(\frac{d}{n}\Bigr)^d \Bigl(1-\frac{d}{n}\Bigr)^{n-d}= A_d\, 2^{-n\,h(d/n)}
\end{equation}
where
\[h(x)=-x\,\log_{2}x-(1-x)\,\log_{2}(1-x)\]
is the binary entropy function.

We can reformulate this to the following well-known sufficient condition for a code to be ugly
(see e.g. \cite[Theorem 2.11]{K} or \cite{Perry}):
\begin{equation}
\label{per}
A_d > 2^{ k-n+n\,h(d/n) }.
\end{equation}

We showed in Example \ref{ex0} that $S_{2^k-2^{k-m-1},k}$ is proper for all $m< k$.
In general, $S_{n,k}$ and $C_{n,n-k}$ may be ugly
for some values of $n$ when $2^k-2^{k-m}< n<2^k-2^{k-m-1}$.

We have checked that $S_{n,k}$ is proper for all $n\ge 2^{k-1}$ when $k\le 8$.
When $k\ge 9$, $S_{n,k}$ is ugly for some values of $n$. An example is $S_{320,9}$
in  Fig. \ref{s320}.

\begin{lemma}
\label{var}
  For a given $k$, let
\[g(n)=k-n+n\,h\Bigl(\frac{d}{n}\Bigr).\]
Then $g(n)$ is increasing with $n$ on
$[2^k-2^{k-m},n(k,m)]$
and decreasing with increasing $n$ on
$[n(k,m),2^k-2^{k-m-1}]$,
where $n(k,m)$ was given in (\ref{nrm}).
\end{lemma}

\begin{IEEEproof}
  From the definitions of $h(x)$ and $g(n)$, we get
\[g(n)=k-n + n\,\log_2 n - d\,\log_2 d - (n-d)\,\log_2 (n-d).\]
By Lemma \ref{dmin}a), $d$ is constant for
$n\in [2^k-2^{k-m},n(k,m)]$.
Considering $n$ as a real variable for the moment, direct calculations gives
\[ \frac{dg(n)}{dn} = -1+\log_2\Bigl(\frac{n}{n-d}\Bigr).\]
Since $d< n/2$, we get $\frac{dg(n)}{dn} < 0$.

Similarly, for $n\in [n(k,m),2^k-2^{k-m-1}]$, $n-d$ is constant by Lemma \ref{dmin}b), and so
\[ \frac{dg(n)}{dn} = -1+\log_2\Bigl(\frac{n}{d}\Bigr)>0.\]
\end{IEEEproof}

Note: The weight distribution of $S_{n(k,m),k}$ was given in Table \ref{tabnr}.
In particular, $A_d=2^m$ for $n=n(k,m)$.
Moreover, $A_d\ge 2^m$
for all $n\in [2^k-2^{k-m},2^k-2^{k-m-1}]$. Hence, (\ref{per}) is satisfied for some
such $n$ if and only if it is satisfied for $n=n(k,m)$.

  For $n=n(k,m)=2^{k-m-2}(2^{m+2}-3)$ we have, from Table \ref{tabnr}, that
\[d=2^{k-m-1}(2^m-1),\]
and so
\[ \frac{d}{n}=\frac{2^{m+1}-2}{2^{m+2}-3}.\]

 For a fixed $m$, let
\begin{equation}
\label{kb0}
G(k)=k-m-2^{k-m-2}U_m
\end{equation}
where we consider $k$ a real variable, and where
\begin{equation}
\label{u-def}
U_m=(2^{m+2}-3)\Bigl\{1-h\Bigl( \frac{2^{m+1}-2}{2^{m+2}-3} \Bigr)\Bigr\}.
\end{equation}

Then (\ref{per}), for $n=n(k,m)$, can be rewritten as
\begin{equation}
\label{kb}
G(k)<0.
\end{equation}

We get
\begin{align}
G'(k)&= 1-2^{k-m-2}U_m \ln 2, \label{kb1}\\
G''(k)&= -2^{k-m-2}U_m (\ln 2)^2. \label{kb2}
\end{align}

To analyze $G(k)$ further, we first give some relations for $h(x)$.

\begin{lemma}
\label{h-lemma}
  For $0< x <1/2$, we have
\begin{equation}
\label{ent}
h(x)< 1- \frac{(1-2x)^2}{2\ln 2}
\end{equation}
and
\begin{equation}
\label{entl}
h(x)> 1- \frac{(1-2x)^2}{2\ln 2} - \frac{\ln 2-\frac{1}{2}}{\ln 2}\cdot (1-2x)^4.
\end{equation}
\end{lemma}

\begin{IEEEproof}
Using Taylor's theorem, we get
\begin{equation}
\label{ht}
h(x)= 1- \frac{1}{\ln 2}\sum_ {i=1}^{\infty} \frac{(1-2x)^{2i}}{2i(2i-1)} .
\end{equation}
for $0\le x \le 1$.

The upper bound (\ref{ent}) follows immediately since
\[h(x)=1- \frac{(1-2x)^2}{2\ln 2} - \frac{1}{\ln 2}\sum_ {i=2}^{\infty} \frac{(1-2x)^{2i}}{2i(2i-1)}.\]
Next, from (\ref{ht}), we get
\[ 0=h(0)=1-\frac{1}{\ln 2}\sum_ {i=1}^{\infty} \frac{1}{2i(2i-1)}.\]
Since $(1-2x)^{2i}\le (1-2x)^{4}$ for $i\ge 2$ and $x\in (0,1/2)$, we get
\begin{align*}
h(x)&> 1-\frac{(1-2x)^2}{2\ln 2} -\frac{1}{\ln 2}\sum_ {i=2}^{\infty} \frac{(1-2x)^{4}}{2i(2i-1)} \\
&= 1-\frac{(1-2x)^2}{2\ln 2} -\frac{(1-2x)^{4}}{\ln 2}\sum_ {i=2}^{\infty} \frac{1}{2i(2i-1)} \\
&= 1-\frac{(1-2x)^2}{2\ln 2} -\frac{(1-2x)^{4}}{\ln 2}\Bigl\{\ln 2-\frac{1}{2}\Bigr\}.
\end{align*}
\end{IEEEproof}

\begin{lemma}
\label{u-lemma}
  For $m\ge 1$, we have
\[ 2^{m+3} U_m \ln 2 =1+u_m,\]
where
\[ \frac{3}{2^{m+2}-3} <u_m < \frac{3}{2^{m+2}-3} +\frac{(2\ln 2 -1)2^{m+2}}{(2^{m+2}-3)^3}. \]
\end{lemma}

\begin{IEEEproof}
We have
\[1-2\,\frac{2^{m+1}-2}{2^{m+2}-3}=\frac{1}{2^{m+2}-3}.\]
Hence, from (\ref{u-def}) and (\ref{ent}) we get
\begin{align*}
2^{m+3} U_m \ln 2 &> 2^{m+3}\ln 2\,(2^{m+2}-3) \frac{1}{2\ln 2} \frac{1}{ (2^{m+2}-3)^2} \\
&= 1+\frac{3}{ 2^{m+2}-3}.
\end{align*}
This proves the lower bound on $u_m$. Similarly, (\ref{u-def}) and (\ref{entl}) imply the upper bound on $u_m$.
\end{IEEEproof}

In particular, (\ref{kb}) and Lemma \ref{u-lemma} imply that $G(m+1)>0$ for all $m\ge 1$. Since $G''(k)<0$ for all $k$ and $G(k)\rightarrow -\infty$ when $k\rightarrow \infty$, we see that $G(k)=0$ has a unique root in $[m+1,\infty)$, we denote it by $\kappa(m)$.
 Further, $G(k)>0$ for $m+1\le k <\kappa(m)$. Also, $G(k)<0$ and $G(k)$ is decreasing
for $k>\kappa$.
\begin{lemma}
\label{not-int}
For $m\ge 1$, $\kappa(m)$ is not an  integer.
\end{lemma}
\begin{IEEEproof}
For $n=n(k,m)=2^{k-m-2}(2^{m+2}-3)$, we have
\[d=2^{k-m-1}(2^m-1),\]
\[n-d=2^{k-m-2}(2^{m+1}-1).\]
and $A_d=2^m$. By definition, $k=\kappa(m)$ if
\[2^{k-n}=2^m\,\Bigl(\frac{d}{n}\Bigr)^d \Bigl(1-\frac{d}{n}\Bigr)^{n-d} \]
or equivalently,
\begin{equation}
\label{eq2}
2^{k-n}\,n^n=2^m\,d^d\, (n-d)^{n-d}.
\end{equation}
Hence, if $k=\kappa(m)$ were an integer, then the exact powers of 2 dividing the two sides of (\ref{eq2}) would be the same.
We will show that this is not the case.

The exact power of 2 dividing $2^{k-n}\,n^n$ is
\[k-n+n(k-m-2).\]

The exact power of 2 dividing $2^m\,d^d\, (n-d)^{n-d}$ is
\begin{align*}
m+d&(k-m-1)+(n-d)(k-m-2)\\
    &= m+d+n(k-m-2)\\
    &> k-n+n(k-m-2),
\end{align*}
that is, we have a contradiction. Hence, $\kappa(m)$ is not an integer.
\end{IEEEproof}

Let $K(m)$ be the smallest integer $k$ such that $G(k)<0$. Then $K(m)>\kappa(m)$ and so we have the following:
\begin{equation}
\label{k01}
K(m)=\lceil \kappa(m)\rceil.
\end{equation}
In Table \ref{k0tab} we give the values of $K(m)$ for $m\le 356$.
\begin{table}[!t]
\caption{The values of $K(m)$ for $1 \le m \le 356$}
\label{k0tab}
\[\begin{array}{|l|cccc|}\hline
m\in      & \{1\} &  [2,4] & [5,14] & [15,36]  \\ \hline
K(m) & 9 & 2m+8 & 2m+9 & 2m+10  \\ \hline \hline
m \in     & & [37,81] & [82,172]  & [173,356] \\ \hline
K(m) & & 2m+11  & 2m+12 & 2m+13 \\ \hline
\end{array}\]
\end{table}

\begin{table*}[!t]
\caption{Values of $n$ for which (\ref{per}) is satisfied, and therefore, both
$S_{n,k}$ and $C_{n,n-k}$ are ugly}
\label{pek-n}
\[\begin{array}{|r|c|c|c|c|}
\hline
k & m=1& m=2& m=3& m=4\\ \hline
9 & [315,324]&&&\\ \hline
10 & [599,676]&&&\\ \hline
11 & [1140,1396]&&&\\ \hline
12 & [2219,2878]& [3286,3367]&&\\ \hline
13 & [4331,5853]& [6458,6844]&&\\ \hline
14 & [8540,11878]& [12717,13888]& [14812,14883]&\\ \hline
15 & [16870,23966]& [25208,28006]& [29371,30013]&\\ \hline
16 & [33486,48290]& [50034,56408]& [58305,58368]\,[58370,60396]\, [60416,60461]& [62460,62468]\\ \hline
17 & [66546,66560]\,[66593,97028] & [99602,113304]& [116102,121434]& [124378,125472] \\ \hline
18 & [132560,194804]& [198432,227423] & [231354,231424]\,[231451,243631]\,[243712,243730]& [247954,251741] \\ \hline
\end{array}\]
\end{table*}
We will next determine good bounds on $\kappa(m)$. These can in turn be used to determine $K(m)$.
We use the notations
\begin{align*}
\lambda&=\log_2(\ln 2) \approx -0.5287663728,\\
\Lambda&=5+\lambda,\\
\mu&=\mu(m)= m+\Lambda,\\
\rho(m)& = m+\mu +\log_2 \mu=2m+5+\log_2(\mu\ln 2) \\
       &= 2m+5+\log_2 \mu+\lambda.
\end{align*}

\begin{lemma}
\label{k1eks}
Let $c>0$.

\noindent a) If
\begin{equation}
\label{bb}
(2^c-1)\mu + 2^c \mu\, u_m > c+\log_2 \mu,
\end{equation}
then
\[\kappa(m)<\rho(m)+c.\]

\noindent b) If
\begin{equation}
\label{bb2}
(2^c-1)\mu + 2^c \mu\, u_m < c+ \log_2 \mu ,
\end{equation}
then
\[\kappa(m)>\rho(m)+c.\]
\end{lemma}

\begin{IEEEproof}
Let $k=\rho(m)+c=m+ \mu +\log_2 \mu+c $. Since $\mu=m+5+\log_2(\ln 2)$, we have
\begin{align*}
G(k) &= k-m-2^{k-m-2}\, U_m \\
&= \mu +\log_2 \mu+c - 2^{m+3+c} \mu \ln 2\, U_m.
\end{align*}
By Lemma \ref{u-lemma},

\[G(k)=\mu +\log_2 \mu+c-2^{c}\mu\, (1+u_m).\]
If (\ref{bb}) is satisfied, then $G(k)<0$ and so $k>\kappa(m)$.
This proves a) and the proof of b) is similar.
\end{IEEEproof}

Let
\begin{align*}
\overline{\omega}(m) &=\rho(m)+ \frac{\log_2 \mu}{\mu \ln 2},\\
\underline{\omega}(m)&=\rho(m)+ \frac{\log_2 \mu}{\mu \ln 2}- \frac{(\log_2 \mu)^2}{ 2\mu^2 \ln 2}.
\end{align*}

\begin{corollary}
\label{k1eksc}
If $m\ge 2$, then
\[\underline{\omega}(m)<\kappa(m)< \overline{\omega}(m).\]
\end{corollary}

\begin{IEEEproof}
 First we consider the upper bound.
Let $c=\frac{\log_2 \mu}{\mu \ln 2}$.
Then
\begin{align*}
(2^c-1)\mu&= (e^{c\ln 2}-1)\mu= \mu \sum_{i=1}^\infty \frac{(\log_2 \mu)^i}{i! \mu ^i} \\
&> \log_2 \mu+ \frac{(\log_2 \mu)^2}{2\mu } \\
&\ge \log_2 \mu+ \frac{\log_2 \mu}{\mu \ln 2}=c+\log_2 \mu
\end{align*}
for $\log_2 \mu\ge \frac{2}{\ln 2}\approx 2.885$, that is, $\mu \ge 7.389$, i.e. $m\ge 3$.
 For $m=2$ we get $(2^c-1)\mu -(c+\log_2 \mu)\approx 0.0468>0$ also. In particular, (\ref{bb}) is satisfied for all $m\ge 2$.
The upper bound therefore follows from Lemma \ref{k1eks}a).

The proof of the lower bound is similar.  For $m=2,3,4$ we can show it by direct computation.
Some calculus shows that $(2^c-1)\mu<\log_2 \mu$ for all $m\ge 2$ and $\mu\, u_m<c$ for
$m\ge 5$. We skip the details. The lower bound therefore follows from Lemma \ref{k1eks}b).
\end{IEEEproof}

 From Corollary \ref{k1eksc} we immediately get the following result.

\begin{corollary}
\label{k0main1}
We have $\kappa(m)-\rho(m)\rightarrow 0$ when $m\rightarrow\infty$.
\end{corollary}

\begin{theorem}
\label{k0main2}
 For all $m\ge 2$, we have
\begin{equation}
\label{k0u}
K(m)= \lceil\underline{\omega}(m)\rceil
\end{equation}
or
\[K(m)= \lceil\underline{\omega}(m)\rceil+1.\]
In particular, if there is no integer between $\underline{\omega}(m)$ and $\overline{\omega}(m)$, then
\[K(m)= \lceil\underline{\omega}(m)\rceil.\]
\end{theorem}

\begin{IEEEproof}
Since $ \overline{\omega}(m) < \underline{\omega}(m) +1$, (\ref{k01}) and Corollary \ref{k1eksc} imply that
\[ \lceil\underline{\omega}(m) \rceil\le K(m)=
\lceil \kappa(m)\rceil\le \lceil\overline{\omega}(m)\rceil
\le \lceil\underline{\omega}(m)\rceil+1.\]
 Further, if there is no integer between $\underline{\omega}(m)$ and $\overline{\omega}(m)$, then
$\lceil\underline{\omega}(m)\rceil =\lceil\overline{\omega}(m)\rceil$.
\end{IEEEproof}

The difference
\[\overline{\omega}(m)- \underline{\omega}(m)= \frac{(\log_2 \mu)^2}{ 2\mu^2 \ln 2} \] is small, except for small $m$. Hence, to have an integer
between $\underline{\omega}(m)$ and $\overline{\omega}(m)$, $\overline{\omega}(m)$ must be close to and above an integer.
If we denote this integer by $2m+5+u$, then $m$ must be close to $\frac{2^u}{\ln 2}-u-5$.
We will make this statement more precise in the following lemma.

\begin{lemma}
\label{theta-n}
Let $u$ be a positive integer.

a) If
\[m\le \frac{2^u}{\ln 2}-u -5- \frac{u^2}{2^u}, \]
then $ \overline{\omega}(m) < 2m+5+u$.

b) If
\[m\ge \frac{2^u}{\ln 2}-u -5+ \frac{u^2}{2^u}, \]
then $ \underline{\omega}(m) > 2m+5+u$.
\end{lemma}

\begin{IEEEproof}
In this proof, we let $m$ be a positive real variable.
We note that $\overline{\omega}(m)$ and $\underline{\omega}(m)$ are still well defined. Moreover, simple
calculus shows that $\frac{d\overline{\omega}(m)}{dm}>0$ and $\frac{d\underline{\omega}(m)}{dm}>0$ for all $m>0$.
Therefore, a) is equivalent to
\[\overline{\omega}\Bigl(\frac{2^u}{\ln 2}-u -5- \frac{u^2}{2^u}\Bigr)<2m+5+u\]
and similarly for b).

Proof of a). Let
$m= \frac{2^u}{\ln 2}-u -5- \frac{u^2}{2^u}$. Then
$\mu= \frac{2^u}{\ln 2}-y$,
where
\[y=u-\lambda+\frac{u^2}{2^u}\] and so
\begin{align*}
\log_2 \mu &= \log_2\Bigl(\frac{2^u}{\ln 2}\Bigr)
 +\log_2\Bigl(1- \frac{y\ln 2}{2^u}\Bigr) \\
 &= u- \lambda
 + \frac{1}{\ln 2} \cdot
\ln\Bigl(1- \frac{y\ln 2}{2^u}\Bigr) \\
&< u- \lambda - \frac{y}{2^u} < u- \lambda.
\end{align*}
Hence
\begin{align*}
\overline{\omega}(m)& = 2m+5+ \lambda+ \log_2 \mu+\frac{\log_2 \mu }{\mu\ln 2} \\
 &< 2m+5+ u- \frac{y}{2^u} + \frac{ u- \lambda}{2^u-y\ln 2 }\\
 & \le  2m+5+ u
\end{align*}
if
\begin{equation}
\label{om-con}
\frac{y}{2^u} \ge \frac{ u- \lambda}{2^u-y\ln 2 }.
\end{equation}
The inequality (\ref{om-con}) is equivalent to
\[\Bigl(u-\lambda+\frac{u^2}{2^u}\Bigr)\Bigl(2^u-y\ln 2\Bigr)\ge 2^u (u-\lambda)\]
which in turn is equivalent to
\[u^2\ge \Bigl(u-\lambda+\frac{u^2}{2^u}\Bigr)^ 2 \ln 2.\]
This is satisfied for all $u\ge 6$.
For $1\le u \le 5$, we can show a) directly by numerical computation. This completes the proof of a).

The proof of b) is similar. We give a sketch, leaving out some details.
Let
$m= \frac{2^u}{\ln 2}-u -5+ \frac{u^2}{2^u}$. Then
$\mu= \frac{2^u}{\ln 2}-y$,
where
\[y=u-\lambda-\frac{u^2}{2^u}\] and so
\begin{align*}
\log_2 \mu &= u- \lambda+ \frac{1}{\ln 2} \cdot
\ln\Bigl(1- \frac{y\ln 2}{2^u}\Bigr) \\
&> u- \lambda - \frac{y}{2^u-y\ln 2}.
\end{align*}
We have
\[\underline{\omega}(m)=2m+5+\lambda+\log_2 \mu+\frac{\log_2 \mu }{\mu\ln 2}- \frac{(\log_2 \mu)^2}{2\mu^ 2\ln 2}.\]
We observe that the function
\[\log_2 x+\frac{\log_2 x}{x\ln 2}- \frac{(\log_2 x)^2}{2x^ 2\ln 2}\]
is increasing with $x$.
Hence,
\begin{align*}
\underline{\omega}(m)&> 2m+5+u-\frac{y}{2^u-y\ln 2}\\
  & \quad +\frac{u- \lambda-\frac{y}{2^u-y\ln 2}  }{2^u-y\ln 2}-\frac{(u- \lambda-\frac{y}{2^u-y\ln 2})^2\ln 2}{2(2^u-y\ln 2)^2} \\
  & \geq 2m+5+u
\end{align*}
if
\[ u- \lambda-y-\frac{y}{2^u-y\ln 2}   \ge \frac{(u- \lambda-\frac{y}{2^u-y\ln 2})^2\ln 2}{2(2^u-y\ln 2)},\]
that is,
\[\frac{u^2}{2^u}\ge \frac{2y+(u- \lambda-\frac{y}{2^u-y\ln 2})^2\ln 2}{2(2^u-y\ln 2)}.\]
This is satisfied for $u\ge 2$. Direct computation shows that b) is true also for $u=1$.
\end{IEEEproof}

Combining Lemmas \ref{theta-n}a) and b), we see that if there is an integer between
$\underline{\omega}(m)$ and $\overline{\omega}(m)$, then this integer is $2m+5+u$ for
 some integer $u$, and $m=m_u$ where
\begin{equation}
\frac{2^u}{\ln 2}-u -5- \frac{u^2}{2^u} < m_u < \frac{2^u}{\ln 2}-u -5+ \frac{u^2}{2^u}.
\label{int}
\end{equation}

We have checked (\ref{int}) for $u\le 10000$.  For $8\le u \le 10000$, there is no integer satisfying (\ref{int}).
For $u\le 7$, there actually is an integer $m_u$ satisfying (\ref{int}). However, in these cases, direct computations
show that $K(m_u)= \lceil\underline{\omega}(m_u)\rceil$. We can therefore conclude that
$K(m)= \lceil\underline{\omega}(m)\rceil$
for $2\le m < 2\cdot 10^{3010}$. This is, of course, far beyond what is needed for any practical application.
Whether there are any $u>10000$ such that there is an integer satisfying (\ref{int}) remains an open question.
However, the length of the interval in (\ref{int}) is $\frac{2u^2}{2^u}$ and
\[ \sum_{u=10001}^\infty \frac{2u^2}{2^{u}} \approx 1.003\cdot 10^{-3002}.\]
Therefore, it is highly unlikely that there is an integer satisfying (\ref{int}) for some $u>10000$. Based on this, we conjecture that $K(m)= \lceil\underline{\omega}(m)\rceil$ for all $m\ge 2$.
\smallskip

If $k\ge K(m)$, we define $b_1(k,m)$ to be the smallest integer and $b_2(k,m)$
the largest integer such that:
\begin{itemize}
\item $b_1(k,m)< n(k,m)<b_2(k,m)$,
\item (\ref{per}) is satisfied for $b_1(k,m) \le n \le b_2(k,m)$.
\end{itemize}
In the next section, we give estimates for $b_1(k,m)$ and $b_2(k,m)$.

 For $k\le 18$, we have computed the values of $n$
in the range $[2^{k-1}+1, 2^k-1]$
for which (\ref{per}) is satisfied. For $k\le 8$, this never happens; and we have checked that $C_{n,n-k}$
is always proper for $k\le 8$.  For $9\le k \le 18$,
the values of $n$ for which (\ref{per}) is satisfied are given in Table \ref{pek-n}. Since $K(5)=19$, we only
have to consider $m\le 4$ when $k\le 18$.

We see that, in general, for any given $k$ and $m$ the set of $n$ where (\ref{per}) is satisfied
consists of zero or more intervals.

\begin{example}
Typically, we have several intervals, except for small values of $k$. We describe $k=17$, $m=1$ as
an example to illustrate why this is the case.
We first give a small list of values in Table \ref{tab17}.
\begin{table}[!t]
\caption{Some values of $A_d$ and $2^{k-n+n\, h(d/n)} $ for $k=17$ and $66545 \le n \le 66593$}
\label{tab17}
\[\begin{array}{|c|cc|}\hline
n & A_d & 2^{k-n+n\, h(d/n)} \\ \hline
66545 & 62 & 62.4 \\
66546 & 62 & 61.5 \\
66560 & 62 & 49.7 \\
66561 & 30 & 48.9 \\
66592 & 30 & 30.3 \\
66593 & 30 & 29.8 \\
\hline
\end{array}\]
\end{table}
We see that (\ref{per}) is satisfied for $n=66593$, but not for $n=66592$. Since
(\ref{per}) turns out to be satisfied for $66593\le n \le 97028$, we get
$b_1(17,1)=66593$. For $n$ in the range $[66561,66593]$ we have $A_d=30$. Since $2^{k-n+n\, h(d/n)}$
is decreasing with increasing $n$, (\ref{per}) is not satisfied for $n$ in the range $[66561,66592]$.
However, we see that for $n=66560$, we have a jump in the value of $A_d$
compared to $n=66561$, and (\ref{per}) is again satisfied for all $n$ in the range $66546\le n \le 66560$.
  For $n=66545$, (\ref{per}) is again not satisfied.
\end{example}

\begin{example}
\label{ex16}
An interesting example occurs when $k=16$ and $m=3$;
$n=58369=b_1(16,3)-1$ is an isolated value of $n$ for which (\ref{per}) is not satisfied.
We have
\[n=2^{16}-2^{13}+2^{10}+1\mbox{ and }d=2^{15}-2^{12}.\]
We have $A_d=8$ and
\[\frac{A_d\, (d/n)^d (1-d/n)^{n-d}}{2^{k-n}}\approx 0.989\]
and so (\ref{per}) is not satisfied. However, $A_{d+1}=16$ and
\[\frac{A_{d+1}\, (d/n)^{d+1} (1-d/n)^{n-d-1}}{2^{k-n}}\approx 1.9106,\]
so the contribution from this term alone is sufficient to conclude that $S_{58369,16}$
is not satisfactory after all. This shows that if (\ref{per}) is not satisfied, but the second
lowest weight of $S_{n,k}$ is close
the minimum weight $d$, it may be a good idea to consider the contribution from this weight also.
\end{example}

  For $k\le 12$ we have checked if the codes $S_{n,k}$ and $C_{n,n-k}$
are proper when (\ref{per}) is not satisfied.
It turns out that this is always the case for $C_{n,n-k}$, but not for $S_{n,k}$.
As an illustration, in Table \ref{tab:Tab3} we give the range of values $n$ for $m=1$ and $9\le k\le 18$, such that
$S_{n,k}$ is proper.
\begin{table}[!t]
\caption{ Ranges of $n$ where $S_{n,k}$ is proper when $2^{k-1}+1\le n\le 2^{k-1}+2^{k-2}$. }
\label{tab:Tab3}
\[\begin{array}{|r|cc|} \hline
k & & \\ \hline
9 & [257,307]   & [331,384] \\ \hline
10 & [513,587]   & [688,768] \\ \hline
11 & [1025,1124] & [1424,1536] \\ \hline
12 & [2049,2195] & [2904,3072] \\ \hline
13 & [4097,4298] & [5908,6144] \\ \hline
14 & [8193,8489] & [11937,12288] \\ \hline
15 & [16385,16798] & [24081,24576] \\ \hline
16 & [32769,33376] & [48422,49152] \\ \hline
17 & [65537,66388] & [97245,98304] \\ \hline
18 & [131073,132321] & [195096,196608] \\ \hline
\end{array}\]
\end{table}


\section{Approximate Analysis
\label{sec:Approximate-Analysis}}

In Table \ref{pek-n} we see that for a given $m$, an increasing fraction
of the codes are ugly
when $k$ increases. Define $\beta_1(k,m)$ and $\beta_2(k,m)$ by
\[ b_1(k,m)=2^k-2^{k-m}+\beta_1(k,m)\]
and
\[ b_2(k,m)=2^k-2^{k-m}+\beta_2(k,m).\]
Clearly, $0< \beta_1(k,m)< \beta_2(k,m)< 2^{k-m-1}$.
Let
\begin{align*}
\gamma_1(k,m)=\, &(k-m)\ln 2 \\
&+ \sqrt{(k-m)^2 (\ln 2)^2+2(k-m)(2^{k}-2^{k-m})\ln 2}.
\end{align*}

\begin{theorem}
\label{ny-con}
We have
\[\beta_1(k,m)\le \lceil \gamma_1(k,m)\rceil.\]
\end{theorem}

\begin{IEEEproof}
Let $n= 2^{k}-2^{k-m}+\sigma$, where $0\le \sigma \le 2^{k-m-2}$.
By Lemma \ref{dmin}, $d=2^{k-1}-2^{k-m-1}$, and by Lemma \ref{Admin},
$A_d\ge 2^m$.
By (\ref{per}) we see that if
\begin{equation}
\label{perapx}
m>k-n+n\,h(d/n),
\end{equation}
then $A_d\ge 2^m >2^{k -n+n\,h(d/n) }$, and so $S_{n,k}$ is ugly.
We have
\[1-2\,\frac{d}{n}=\frac{n-2d}{n}= \frac{\sigma}{n} .\]
By (\ref{ent}),
\[h\Bigl(\frac{d}{n}\Bigr) < 1- \frac{1}{2\ln 2}\cdot \frac{\sigma^2}{n^2}.\]
Hence
\begin{align*}
k -n+n\,h\Bigl(\frac{d}{n}\Bigr) &< k-\frac{\sigma^2 }{2n\ln 2}\\
&= k- \frac{\sigma^2 }{2(2^{k}-2^{k-m}+\sigma)\ln 2}.
\end{align*}
By (\ref{perapx}), if
\begin{equation}
\label{par2}
m\ge k- \frac{\sigma^2 }{2(2^{k}-2^{k-m}+\sigma)\ln 2},
\end{equation}
then $S_{n,k}$ is ugly.
Since (\ref{par2}) is equivalent to $\sigma\ge \gamma_1(k,m)$, and $\sigma$ is an integer, that is
$\sigma\ge \lceil \gamma_1(k,m)\rceil$,
we can conclude that $ \beta_1(k,m)\le \lceil \gamma_1(k,m)\rceil$.
This proves the theorem.
\end{IEEEproof}

 \emph{Remark}. We see that $\beta_1(k,m)-1$ is an upper bound on the number of
$n$ in $[2^k-2^{k-m}+1,2^k-2^{k-m}+2^{k-m-2}]$ such that $S_{n,k}$ is satisfactory.
Therefore, a main corollary of Theorem \ref{ny-con} is that, for any fixed $m$,
$\beta_1(k,m)/2^{k-m-2}$ converges to 0 exponentially fast when $k$ increases.

  For $2^{k}-2^{k-m}+2^{k-m-2}< n< 2^{k}-2^{k-m-1}$, we have a similar result.
Let
\begin{align*}
\lefteqn{\gamma_2(k,m)=-(k-m)\ln 2} \\
&\qquad + \sqrt{(k-m)^2 (\ln 2)^2+2(k-m)(2^{k}-2^{k-m-1})\ln 2}.
\end{align*}

\begin{theorem}
\label{ny-con2}
We have
\[\beta_2(k,m)\ge 2^{k-m-1}-\lceil \gamma_2(k,m)\rceil.\]
\end{theorem}

\begin{IEEEproof}
The proof is similar to the proof of Theorem \ref{ny-con}. Let
\[n= 2^{k} -2^{k-m-1}- \zeta ,\]
where $0<\zeta< 2^{k-m-2}$. From Lemma \ref{dmin} we get
\[d=2^{k-1} -2^{k-m-2}- \zeta ,\]
and so
\[1-2\,\frac{d}{n}=\frac{n-2d}{n}= \frac{\zeta}{n} .\]
Therefore, analogously to (\ref{par2}) we get
\[m\ge k- \frac{\zeta^2}{2(2^k-2^{k-m-1}-\zeta)\ln 2}\]
is a sufficient condition for $S_{n,k}$ to be ugly.
Since this is equivalent to $\zeta\ge \lceil\gamma_2(k,m)\rceil$, Theorem \ref{ny-con2} follows.
\end{IEEEproof}

\begin{theorem}
\label{cor}
Let $N_k$ be the number of $n \in [2^{k-1},2^k-1]$ such that $S_{n,k}$ is satisfactory.
Then
\[N_k < k+\frac{2^{(k+5)/2}}{3}\sqrt{k^3\,\ln 2}.\]
\end{theorem}

\begin{IEEEproof}
The number $N_{k,m}$ of satisfactory codes for $n$ in the interval $[2^k-2^{k-m}+1,2^{k}-2^{k-m-1}]$ is at most
those for $n\in [2^k-2^{k-m}+1,b_1(k,m)-1]$ plus
those for $n\in [b_2(k,m)+1,2^{k}-2^{k-m-1}]$.
The number of $n$ in the first interval is
\begin{align*}
b_1(k,m)-1-(2^k-2^{k-m}) &= \beta_1(k,m)-1 \\
& \le \lceil \gamma_1(k,m)\rceil -1< \gamma_1(k,m).
\end{align*}
The number of $n$ in the second interval is
\begin{align*}
2^{k}-2^{k-m-1}-b_2(k,m) &=2^{k-m-1}-\beta_2(k,m) \\
& \le \lceil\gamma_2(k,m)\rceil<\gamma_2(k,m)+1.
\end{align*}
We see that, for $1\le m \le k-1$, we have
\begin{align*}
(k-m)\ln 2&\,+2(2^k-2^{k-m}) \\
&< (k-m)\ln 2+2(2^k-2^{k-m-1})
< 2^{k+1}.
\end{align*}
Hence
\[N_{k,m}< \gamma_1(k,m)+\gamma_2(k,m)+1< 1+2\sqrt{(k-m)2^{k+1}\ln 2},\]
and so
\begin{align*}
N_k &= 1+\sum_{m=1}^{k-1} N_{k,m} \\
&< 1+\sum_{m=1}^{k-1} \bigl(1+2 \sqrt{(k-m)2^{k+1}\ln 2}\bigr) \\
&= k+2\sum_{m=1}^{k-1}\sqrt{(k-m)2^{k+1}\ln 2} \\
&=k+ 2^{(k+3)/2}\sqrt{\ln 2} \sum_{m=1}^{k-1} \sqrt{k-m} \\
&<k+ 2^{(k+3)/2}\sqrt{\ln 2}\cdot \frac{2}{3} \sqrt{k^3}.
\end{align*}
\end{IEEEproof}

\begin{corollary}
\[\frac{ N_k}{2^{k-1}}\rightarrow 0,\]
when $k\rightarrow \infty$.
\end{corollary}

Theorem \ref{cor} shows that $C_{n,n-k}$ is ugly for most values of $n$.
On the other hand, $C_{n,n-k}$ is satisfactory for many values of $n$ as shown by Corollary \ref{c-kange}.


\section{A generalization of the construction
\label{sec:Generalization}}

The matrix $H_k$ was defined by concatenating $H_k^{(k-m)}$ for $m=1,2,\ldots , k$.
We can generalize this by concatenating $t_1$ copies of $H_k^{(k-1)}$, followed
by $t_2$ copies of $H_k^{(k-2)}$, $t_3$ copies of $H_k^{(k-3)}$, etc. for any
sequence $(t_1,t_2,\ldots ,t_k)$ of positive integers. Most of the previous results
carries over, with obvious modifications.  For now, we only consider the construction with $t_i=1$ for
$i\ge 2$, and we write $t_1=t$. As before, we use the notation $S_{n,k}$ for the codes generated by the first
$n$ columns of the matrix.  For large $t$, these codes have low rate. The dual codes
will have very high rate and minimum distance 2.

Consider $S_{n',k}$ generated by $H_k(n')$, and let $S_{n,k}$, where
\begin{equation}
\label{tn}
n=2^{k-1}(t-1)+n'\in [2^{k-1}t,2^{k-1}(t+1)-1],
\end{equation}
be the code generated by the matrix
\[ H_k(n)=\overbrace{ H_k^{(k-1)}| H_k^{(k-1)}| \cdots |H_k^{(k-1)}}^{t-1}|H_k(n').\]

We see that we get a code $S_{n,k}$ for each $n\ge 2^{k-1}$.
Also, given $n$, the values of $t$ and $n'$ are uniquely determined by (\ref{tn}).

 From its definition, we immediately get the following lemma.

\begin{lemma}
\label{lenging}
a) The weight of the first row of $H_k(n)$ is $2^{k-1}(t-1)$ larger than the weight of
the first row of $H_k(n')$.

b) For any other non-zero codeword in $S_{n,k}$, the weight is $2^{k-2}(t-1)$
larger than the weight of the
corresponding codeword in $H_k(n')$.
\end{lemma}
In particular, we see that
\begin{itemize}
\item The minimum distance of $S_{n,k}$ is $2^{k-2}(t-1)$ larger than the minimum distance of $H_k(n')$.
\item For a non-zero codeword of $S_{n,k}$ of weight $w$,
either $w\ge n/2$ or there is a unique other codeword in the code of weight $n-w$.
\end{itemize}
This last property was used to prove Theorem \ref{S-proper}.
Therefore, this theorem can be directly generalized by a similar proof.
Let
\[ \overline{\tau}_{t,k,m} =\min \Bigl\{2^{k-m-2},\frac{1+\sqrt{1+2^{k+1}(t+1)-2^{k-m+2}} }{2}\Bigr\},\]
\begin{align*}
\underline{\tau}_{t,k,m} & =\min \Bigl\{2^{k-m-2}-1,\\
& \qquad \qquad \frac{-1+\sqrt{1+2^{k+1}(t+1)-2^{k-m+1}} }{2}\Bigr\}.
\end{align*}

Note that $\overline{\tau}_{1,k,m}=\overline{\tau}_{k,m}$ and $\underline{\tau}_{1,k,m} =\underline{\tau}_{k,m} $.

\begin{theorem}
\label{S-propek-t}
 For $t\ge 1$ and $k>m\ge 1$, if
\begin{equation}
\label{tcon1}
2^{k-1}(t+1)-2^{k-m} \le n \le 2^{k-1}(t+1)-2^{k-m}+ \overline{\tau}_{t,k,m}
\end{equation}
or
\begin{equation}
\label{tcon2}
2^{k-1}(t+1)-2^{k-m-1}-\underline{\tau}_{t,k,m} \le n \le 2^{k-1}(t+1)-2^{k-m-1},
\end{equation}
then $S_{n,k}$ is proper.
\end{theorem}

The proof is similar to the proof of Theorem \ref{S-proper} and is omitted.

\begin{theorem}
\label{tp-kange}
a) If $m\ge 1$ and
\begin{equation}
\label{tmb}
m\ge \Bigl\lceil \frac{k-3-\log_2t}{2}\Bigr\rceil,
\end{equation}
then $S_{n,k}$ is proper for all
\[n\in[2^{k-1}(t+1)-2^{k-m},2^{k-1}(t+1)-2^{k-m-1}].\]

b) $S_{n,k}$ is proper for all
\[n\in \Bigl[2^{k-1}(t+1)- 2^{k- \bigl\lceil \frac{k-3-\log_2t}{2}\bigr\rceil },2^{k-1}(t+1)-1\Bigr].\]
\end{theorem}

\begin{IEEEproof}
Similarly to Theorem \ref{p-kange}, we see that if
\begin{equation}
\label{tx}
x^2+6\, x \le 2^{k+1}(t+1),
\end{equation}
where $x=2^{k-m-1}$,
then $S_{n,k}$ is proper for all
\[n\in [2^{k-1} (t+1)-2^{k-m},2^{k-1} (t+1)-2^{k-m-1}].\]
We see that if $2^{k-2\,m-3}\ge t+1$, then $x^2\ge 2^{k+1}(t+1)$ and so (\ref{tx})
is not satisfied.
However, if
\begin{equation}
\label{txx}
2^{k-2\,m-3}\le t,
\end{equation}
then $x^2\le 2^{k+1}t$ and so
\[x^2+6\, x \le 2^{k+1}t+6\cdot 2^{k-m-1}\le 2^{k+1}(t+1)\]
for $6\cdot 2^{-m-2}\le 1$, that is, all $m\ge 1$.
Since (\ref{txx}) is equivalent to $k-2\,m-3\le \log_2 t$, we get the theorem.
\end{IEEEproof}

\begin{theorem}
\label{skr}
For $k\ge 6$ there exists an integer $\theta(k)\le 2^{k-5}$ such that $S_{n,k}$ is proper for all
$n\ge 2^{k-1}\,\theta(k)$.
\end{theorem}

\begin{IEEEproof}
 For $k\ge 6$ and $t\ge 2^{k-5}$, we get
\[\frac{k-3-\log_2t}{2}\le \frac{k-3-(k-5)}{2}=1.\]
By Theorem \ref{tp-kange}b),
$S_{n,k}$ is proper for all
\[n\in [2^{k-1}(t+1)-2^{k-1},2^{k-1}(t+1)-1].\]
This implies that $S_{n,k}$ is proper for all 

 $n\ge 2^{k-1}\cdot 2^{k-5}=2^{2k-6}.$
\end{IEEEproof}

\begin{corollary}
\label{skr1}
If $k\ge 6$, then $S_{n,k}$ is proper for all
\[n\ge 2^{2k-6}-3\cdot 2^{k-3}+2.\]
\end{corollary}

\begin{IEEEproof}
By Theorem \ref{skr}, $S_{n,k}$ is proper for all
$n\ge 2^{2k-6}$.
Next, Theorem \ref{tp-kange}b) for $t=2^{k-5}-1$
shows that $S_{n,k}$ is proper for
\[n\in [2^{2k-6}-2^{k-2},2^{2k-6}-1].\]

 Finally, Theorem \ref{S-propek-t} for $t=2^{k-5}-1$, $k\ge 6$, and $m=1$
implies that $S_{n,k}$ is proper for
\[n\in [2^{2k-6}-2^{k-2}-\lfloor\underline{\tau}_{2^{k-5}-1,k,1}\rfloor,2^{2k-6}-2^{k-2}].\]
It remains to show that
\begin{equation}
\label{tau-verdi}
\lfloor\underline{\tau}_{2^{k-5}-1,k,1}\rfloor=2^{k-3}-2.
\end{equation}
We have
\begin{align*}
\underline{\tau}_{2^{k-5}-1,k,1} &=\frac{-1+\sqrt{1+2^{2k-4}-2^k}}{2} \\
&=\frac{-1+\sqrt{(2^{k-2}-2)^2-3}}{2}\\
&=\frac{-1+\sqrt{(2^{k-2}-3)^2+2^{k-1}-8}}{2},
\end{align*}
and so, for $k\ge 6$,
\[\frac{-1+(2^{k-2}-3)}{2}< \underline{\tau}_{2^{k-5}-1,k,1}<\frac{-1+(2^{k-2}-2)}{2}.\]
This proves (\ref{tau-verdi}).
\end{IEEEproof}
\smallskip

Until recently, the best general result of this kind was \cite[Theorem 2.64]{K}:
If $k\ge 5$ and
\[n\ge (2^k-1)(2^k-3),\]
then there exists a proper $[n,k]$ code.

This bound was recently improved in \cite{KY} to the following:
If $k\ge 5$ and
\[n\ge 2^{k-1} \bigl( 2^{k-5} + 2^{\lfloor (k-5)/2\rfloor} \bigr),\]
then there exists a proper (and self complementary) $[n,k]$ code.

Clearly, Theorem \ref{skr} above gives a further improvement and is now the best known
such bound.

We can also find lower bounds on $\theta(k)$. We consider $S_{n,k}$ for $n$ in the middle of the interval
with $m=1$, that is 
\[n=n(k,1)+2^{k-1}(t-1)=2^{k-3}(4t+1),\]
 where $n(k,1)$ was defined in (\ref{nrm}).
 When we consider only the term in $S_{n,k}$ of lowest degree, we know that
the case $n=n(k,1)+2^{k-1}(t-1)$ is the worst case (cfr. Lemma \ref{var}). 
Moreover, this term of lowest degree is the dominating
one in $S_{n,k}$. Therefore, it is reasonable to consider these values of $n$ when we look for non-proper $S_{n,k}$.

\begin{theorem}
\label{t-low}
For $k\ge 6$ we have
\begin{align}
\mbox{a) }&\qquad \theta(k)\ge \theta_1(k)=\min\{t \mid S_{2^{k-3}(4t+1),k}\mbox{ is proper}\}, \label{bound1}\\
\mbox{b) }&\qquad \theta(k)\ge \theta_2(k)=\Bigl\lceil \frac{2^{k-6}}{(k-1)\ln 2}-\frac{1}{4} \Bigr\rceil. \label{bound2}
\end{align}
\end{theorem}

\begin{IEEEproof}
Proof a). We see that if $S_{2^{k-3}(4t+1),k}$ is not proper,
then by the definition of $\theta(k)$, $\theta(k)>t$. Therefore, (\ref{bound1}) follows.

Proof b).
Again, consider $S_{n,k}$ for $n=2^{k-3}(4t+1)$.
Then, by Table \ref{tabnr} and Lemma \ref{lenging},
\[d=2^{k-2}+2^{k-2}(t-1) =2^{k-2}t.\]
Since $A_d=2$, (\ref{per}) implies that the code is ugly if
\begin{equation}
\label{t-con}
1>k-n+n\,h\left(\frac{d}{n}\right).
\end{equation}
We have
\[\frac{d}{n}=\frac{2t}{4t+1} \mbox{ and } 1-2\frac{d}{n}=\frac{1}{4t+1}, \]
and so, by (\ref{ent}),
\[-n+n h\Bigl(\frac{d}{n}\Bigr)< n\left(-\frac{\left(\frac{1}{4t+1} \right)^2}{2\ln 2} \right)= - \frac{2^{k-4}}{(4t+1)\ln 2}.\]
Hence, if
\[1\ge k-\frac{2^{k-4}}{(4t+1)\ln 2},\]
that is
\[t\le \frac{2^{k-6}}{(k-1)\ln 2}-\frac{1}{4},\]
then $S_{n,k}$ is ugly. Therefore, $\theta(k)>t$, and the theorem follows.
\end{IEEEproof}

We now give a lemma that is useful for studying when $S_{n,k}$ codes in general are proper for a given $k$.

\begin{lemma}
\label{small-n2}
Let $n\ge 2^{k-1}+1$ and let $w$ be the weight of the first row of $H_k(n)$.
If $P_{\rm ue}(S_{n,k},p)-p^{w}(1-p)^{n-w}$ is increasing on $[0,1/2]$,
then $S_{n+2^{k-1}u,k}$ is proper for all integers $u\ge 0$.
\end{lemma}

\begin{IEEEproof}
The weight of the first row of $H_k(n+2^{k-1}u)$ is $2^{k-1}u+w$.
The weight of any other non-zero codeword in $S_{n+2^{k-1}u,k}$ is $2^{k-2}u$ larger than the
corresponding codeword in $S_{n,k}$. Hence
\begin{align*}
P_{\rm ue}(& S_{n+2^{k-1}u,k},p) \\
    =&\, p^{2^{k-1}u+w}(1-p)^{n-w} \\
     &+ p^{2^{k-2}u}(1-p)^{2^{k-2}u}\Bigl\{  P_{\rm ue}(S_{n,k},p)-p^{w}(1-p)^{n-w} \Bigr\}.
\end{align*}
By assumption, $P_{\rm ue}(S_{n,k},p)-p^{w}(1-p)^{n-w}$ is increasing on $[0,1/2]$.
Since $p^{2^{k-2}u}(1-p)^{2^{k-2}u}$ and $p^{w}(1-p)^{n-w}$ are increasing on $[0,1/2]$,
we can conclude that $P_{\rm ue}(S_{n+2^{k-1}u,k},p)$ is increasing on $[0,1/2]$,
that is, $S_{n+2^{k-1}u,k}$ is proper.
\end{IEEEproof}

For the use of this lemma, it is useful to observe that the conclusion of Theorem \ref{S-propek-t} can be improved:
if (\ref{tcon1}) or (\ref{tcon2}) hold, then $P_{\rm ue}(S_{n,k},p)-p^{w}(1-p)^{n-w}$ is increasing on $[0,1/2]$.
The proof carries over immediately.

Using Lemma \ref{small-n2} and computations, we have determined $\theta(k)$ for $6\le k\le 17$. These values are given in
Table \ref{t0tab} together with the lower and upper bounds on $        \theta(k)$ in Theorems \ref{t-low} and \ref{skr}.
We have also included the bounds for $18\le k\le 20$.
We see that the upper bound is very loose, but $\theta(k)$ equals
the implicit lower bound $\theta_2(k)$  for all $k\le 17$. We conjecture that this may be the case for all $k$.
Table \ref{t0tab} shows that the explicit lower bound $\theta_2(k)$ is also loose (but substantially better
than the upper bound) and the ratio $\theta_1(k)/\theta_2(k)$ is
increasing slowly with $k$. For $k=13$ the ratio is 1.375, for $k=16$ it is 1.485, and for $k=20$ it is 1.555.
Theorem \ref{tilde} below shows that the ratio is always less than 2.
The lower bound $\theta_2(k)$ has the advantage that it is explicit and that it shows that  $\theta(k)$ grows exponentially with $k$.
In Appendix 1 we prove the following theorem.

\begin{theorem}
\label{tilde}
Let $k\ge 6$ and $R=2^{k-4}$.
Let ${{\vartheta}}={\vartheta}(k)$ be the positive real number defined by
\begin{equation}
\label{tildef}
\frac{2(4R-2)(4{\vartheta}+1)\Bigl(R+\sqrt{(R-1)^2-8R{\vartheta}} \Bigr)}{4{\vartheta}R +R-1-(4{\vartheta}+1)\sqrt{(R-1)^2-8R{\vartheta}}}\pi^R = 1,
\end{equation}
where
\begin{equation}
\label{pidef}
\pi=\frac{8R{\vartheta}+R-1- \sqrt{(R-1)^2-8R{\vartheta}}}{8R{\vartheta}+3R-1+ \sqrt{(R-1)^2-8R{\vartheta}}}.
\end{equation}
Then
\begin{align*}
\mbox{a)} & \quad \theta_1(k)\le \lceil {\vartheta}(k) \rceil, \\
\mbox{b)} & \quad {\vartheta}(k) \le \frac{2^{k-5}}{(k-2)\ln 2 + \ln(k-3)-1/(2^{k-3}-1)}+ \frac{1}{2},\\
\mbox{c)} & \quad \theta_1(k)> {\vartheta}(k)- \frac{k-2}{k-3} \bigl(4\vartheta(k) +1\bigr) 4^{-k}.
\end{align*}
\end{theorem}

Combining Theorem \ref{tilde}a) and Theorem \ref{tilde}b), we get the following corollary.
\begin{corollary}
We have
\[\theta_1(k) =\lceil {\vartheta}(k) \rceil\mbox{ or } \theta_1(k) =\lceil {\vartheta}(k) \rceil-1.\]
Moreover, the first alternative is the most likely.
\end{corollary}

We have included $\lceil {\vartheta}(k)\rceil$ in Table \ref{t0tab}.
For the range of values we have computed, i.e. $k\le 20$, we have
$\theta_1(k) =\lceil {\vartheta}(k) \rceil$. If the conjecture
that $\theta(k)=\theta_1(k)$ is true, then $\lceil {\vartheta}(k)\rceil$ is a sharp upper bound on $\theta(k)$.
Further, if the conjecture  is true, then $S_{n,k}$ is proper for all $n\gtrsim 2^{2k-6}/(k\ln 2)$,
a substantially stronger result than what we have been able to show in Corollary  \ref{skr1}.

\begin{table}[!t]
\caption{ Values of and bounds on $        \theta(k)$. }
\label{t0tab}
\[\begin{array}{|l|rrrrrrrrr|} \hline
       k                          & 6 & 7 & 8 &  9 & 10 & 11 &  12  & 13 & 14 \\ \hline
\mbox{Lower bound }\theta_2(k)    & 1 & 1 & 1 &  2 &  3 &  5 &   9  & 16 & 29 \\
\mbox{Lower bound }\theta_1(k)    & 1 & 1 & 1 &  2 &  4 &  7 &  12  & 22 & 41 \\
\theta(k)                         & 1 & 1 & 1 &  2 &  4 &  7 &  12  & 22 & 41  \\
\lceil {\vartheta}(k)\rceil                 & 1 & 1 & 1 &  2 &  4 &  7 &  12  & 22 & 41  \\
\mbox{Upper bound, Thm. }\ref{skr}& 2 & 4 & 8 & 16 & 32 & 64 & 128  & 256 & 512\\ \hline
\end{array}\]
\[\begin{array}{|l|rrrrrr|} \hline
       k                           &   15 &   16 &   17  &   18 &   19  &    20  \\ \hline
\mbox{Lower bound }\theta_2(k)     &   53 &   99 &  185  &  348 &  657  &  1244  \\
\mbox{Lower bound }\theta_1(k)     &   78 &  147 &  279  &  530 & 1012  &  1935  \\
\theta(k)                          &   78 &  147 &  279  &   &   &    \\
\lceil {\vartheta}(k)\rceil                  &   78 &  147 &  279  &  530 & 1012  &  1935  \\
\mbox{Upper bound, Thm. }\ref{skr} & 1024 & 2048 & 4096  & 8192 & 16384 & 32768 \\ \hline
\end{array}\]
\end{table}

For $k\ge 6$, let $\Phi_k$ be the number of $n\in[2^{k-1}+1, 2^{2k-6}-1]$ such that $S_{n,k}$ is proper.

We have shown by direct computation and the use of Lemma \ref{small-n2} that for $6\le k \le 8$, $S_{n,k}$ is proper
for all $n\in [2^{k-1}+1,2\cdot 2^{k-1}]$.
Hence, $\Phi_k=2^{2k-6}-2^{k-1}-1$ for $6\le k \le 8$.

For $9\le k\le 12$ we know that there are values of $n$ where $S_{n,k}$ is not proper.
For given $t$, $k$, $m$, let $X_{t,k,m}$ be the set of
$n\in [2^{k-1}(t+1)-2^{k-m},2^{k-1}(t+1)-2^{k-m-1}]$ for which $S_{n,k}$ is not proper.
The set $X_{t,k,m}$ may be empty. In particular, Theorem
\ref{tp-kange}a) shows that $X_{t,k,m}=\emptyset$ if (38) is satisfied.
On the other hand, Table \ref{tab:Tab3} shows that $X_{1,k,1}\ne\emptyset$ for $9\le k \le 18$.
By direct computation
and the use of Lemma \ref{small-n2}, we have shown that the values given in Table \ref{noprop} are the only values
$n\ge 2^{k-1}+1$ for which $S_{n,k}$  is not proper. The computations have been extended up to $k=17$.
In general, the values in $X_{t,k,m}$ are not necessarily consecutive. For example
\[X_{11,14,1}=[91124,93142]\cup [93181,93184].\]
This is similar to what we have in Table \ref{pek-n}, and the underlying reason is the same.

\begin{table}[!t]
\caption{ Ranges of $n\ge 2^{k-1}+1$ where $S_{n,k}$ is not proper. }
\label{noprop}
\[\begin{array}{|r |r |l|} \hline
k & m & X_{t,k,m},\, 1\le t \le \theta(k)-1 \\ \hline
9 & 1 &[308,330] \\ \hline
10 & 1 & [588,687],\,[1127,1175],\,[1661,1667] \\ \hline
11 & 1 & [1125,1423] ,\,[2200,2402],\,[3255,3397],\\
&  & [4306,4396],\,[5353,5398],\,[6399,6401] \\
 & 2 & [1661,1667] \\ \hline
12 & 1 & [2196,2903],\,[4301,4902],\,[6393,6893], \\
 & & [8500,8901],\, [10582,10917],\, [12661, 12935],  \\
 &   & [14738, 14955],\,[16813, 16977],\,[18887, 19000], \\
   &   & [20959, 21024],\,[23030, 23050] \\
 & 2 & [3255,3397],\,[5353,5398] \\
\hline
\end{array}\]
\end{table}

The conjecture that $\theta(k)=\theta_2(k )$ can be reformulated as follows: if
\[ \bigcup_{t=1}^\infty \bigcup_{m=1}^{k-1} X_{t,k,m}\ne\emptyset, \]
then $n(k,1)\in X_{1,k,1}$. In general, we conjecture that if $X_{t,k,m}\ne\emptyset$, then $2^{k-1}(t-1)+n(k,m)\in X_{t,k,m}$.

 From Table \ref{noprop}  we get the explicit values of
$\Phi_k$ given in Table \ref{theta-tab}. We have included in the table the lower bound given in Theorem \ref{del} below.

\begin{table}[!t]
\caption{ The values of $\Phi_k$ for small $k$. }
\label{theta-tab}
\[\begin{array}{|l|rrrrrrr|} \hline
       k            & 6 & 7 & 8 & 9 & 10 & 11 & 12                                   \\ \hline
\Phi_k            & 31 & 191 & 895 & 3816 & 15715 & 63719 & 256544           \\
{\rm Bound, Thm. \ref{del}} & 16 & 134 & 683 & 3023 & 12677 & 51880 & 209866 \\ \hline
\end{array}\]
\end{table}

\begin{theorem}
\label{del}
When $k\ge 6$, $S_{n,k}$ is proper for
\[\Phi_k\ge \left\lceil \frac{17}{21}\cdot 2^{2k-6}-\frac{55}{3}\cdot2^{k-5} \right\rceil\]
of the values of $n\in[2^{k-1}+1, 2^{2k-6}-1]$.
\end{theorem}
The proof is given in Appendix 2.

\emph{Comment.} We clearly have $\Phi_k\ge 2^{k-1}(2^{k-5}-\theta(k))$. The discussion on $\theta(k)$ above indicates that we may have
$\theta(k)/2^{k-5}\rightarrow 0$.
If this is the case, then we have $\lim_{k\rightarrow \infty} \Phi_k/2^{2k-6}=1$.


\section{Summary and future work
\label{sec:Results}}

In this paper we have analyzed the codes $S_{n,k}$ and their duals $C_{n,n-k}$ (which are shortened Hamming codes)  
to investigate if they are proper or satisfactory codes for error detection.

We have determined the weight distribution of the codes $S_{n,k}$
and computed the undetected error probability.

  For $n=2^k-2^m$, the codes $S_{n,k}$ are proper for all
$1\le m \le k-1$. However, for $k$ greater than 8, there are values of $n$ such that
$S_{n,k}$ and $C_{n,n-k}$ are ugly (not satisfactory)
for error detection. For increasing $k$, the percentage of such codes is increasing.
On the other hand, we have shown that the number values $n$ such that $S_{n,k}$  is proper grows exponentially with $k$.

We have given a generalization of the construction which defines codes $S_{n,k}$ for all lengths greater than $2^{k-1}$, 
and we have shown that $S_{n,k}$ is proper for all $n\ge 2^{2k-6}-3\cdot 2^{k-3}+2$.
 Moreover, $S_{n,k}$ is proper for at least $17/21$ of the shorter lengths.
A plausible conjecture ($\theta(k)=\theta_1(k)$) implies that $S_{n,k}$ probably is proper for all $n\ge 2^{2k-6}/(k \ln 2)$.

An open question for future work is the following: is it the case that if there is an
$n\in[2^{k-1}t, 2^{k-1}(t+1)-1]$ for which $S_{n,k}$ is not proper,
then $ n=2^{k-3}(4t+1)$ is such an $n$?
If the answer is yes (which we believe it is), then this would in particular imply the conjecture referred to above.

 Further work may concern searching for modifications of the construction that will extend the range of lengths where the codes
are proper or satisfactory. In particular, one line of investigation could be to consider lengths less than $2^{k-1}$ by looking
at the the duals of the best known codes of minimum distance 4.


\section*{Appendix 1\\ Proof of Theorem \ref{tilde}}

Let \[P_t(p)=P_{\rm ue}(S_{2^{k-1}t+2^{k-3},k},p).\]
The expression for $P_t(p)$ that was used to determine the
bound $\theta_1(k)$ is easily obtained by combining Table \ref{tabnr} (with $m=1$)
and Lemma \ref{lenging}. Since $R=2^{k-4}$, we have:

\begin{align}
P_t(p) &= 2\,p^{4Rt}(1-p)^{4Rt+2R} \nonumber  + (16R-8)\, [p(1-p)]^{4Rt+R}  \nonumber \\
         & + 4\,p^{4Rt+2R}(1-p)^{4Rt}   + p^{8Rt}(1-p)^{2R}. \label{exp}
\end{align}

We note that $P_t(p)$ is well defined for any positive real number $t$.
 For large values of $t$, $P_t(p)$
is increasing on $[0,1/2]$. For small values of $t$, $P_t(p)$ is first increasing, then decreasing, then again increasing.
There is a limiting $t$ such that, for this $t$, there is a $p_0$ such that $P_t'(p_0)=0$ and $P_t'(p)>0$
for all other $p$ in $(0,1/2)$.
In particular, this implies that $P_t''(p_0)=0$. In principle, the two equations $P_t'(p_0)=0$ and $P_t''(p_0)=0$
can be used to determine $p_0$ and $t=\theta_0$. However, the equations are complicated, and we will consider an approximation
which is easier to handle. We remark at this point that for $t$ close to $R$,  $P_t''(p_0)=0$ is not possible.

In this appendix we often drop $k$ from $\theta(k)$ and write just $\theta$ when the value of $k$ should be clear from the context.
Similarly we write $\vartheta$ for $\vartheta(k)$, $\theta_0$ for $\theta_0(k)$, etc.

In $P_t(p)$, the last three terms are increasing on $[0,1/2]$ whereas
the first term is increasing on $[0,d/n]$ and decreasing on $[d/n,1/2]$.
For all $p\in [0,1/2]$, the first two terms are dominating.
Therefore, we first consider the sum of these two terms:
\[f_t(p)= 2\,p^{4Rt}(1-p)^{4Rt+2R} + (16R-8)\, [p(1-p)]^{4Rt+R}\]
and determine the ${\vartheta}$ and $p_1$ such that $f'_{{\vartheta}}(p_1)=f''_{{\vartheta}}(p_1)=0$.
We expect ${\vartheta}$ to be a good approximation to $\theta_0$ (and $p_1$ to be a good approximation to $p_0$).

The remaining two terms in $P_{{\vartheta}}(p)$ are much smaller. Moreover, they are increasing on $[0,1/2]$. Therefore,
${\vartheta}\ge \theta_0$. In particular, $\theta_1=\lceil \theta_0 \rceil \le \lceil {\vartheta} \rceil$.
This proves Theorem \ref{tilde}a).

We have
\begin{align}
f_t'(p) &= 8Rt\,p^{4Rt-1}(1-p)^{4Rt+2R} \nonumber \\
      &\quad -(8Rt+4R)p^{4Rt}(1-p)^{4Rt+2R-1} \nonumber \\
      &\quad +(16R-8)(4Rt+R)[p(1-p)]^{4Rt+R-1}(1-2p) \nonumber \\
      &= 4R\,p^{4Rt-1}(1-p)^{4Rt+R-1}  \nonumber \\
      &\quad \cdot \Bigl\{[2t-(4t+1)p](1-p)^R \nonumber \\
      &\qquad \quad +(4R-2)(4t+1)(1-2p) p^R \Bigr\}.\label{fdir}
\end{align}

\emph{Remark.} We see that $f_t'(1/2)= -4R\, (1/2)^{8Rt+2R-1}<0$ for all $t$. Hence, the equation $f_{{\vartheta}}'(p)=0$
will, in addition to $p_1$, have at least one more solution in $(0,1/2)$. This second solution will be closer to $1/2$.
However, it does not reflect a property of $P_{{\vartheta}}(p)$, but only the approximation $f_{{\vartheta}}(p)$.

Since $f'_{{\vartheta}}(p)=0$ and $0<p<1/2$, we have
\begin{equation}
\label{equation1}
A\,(1-p)^R+B\,p^R=0
\end{equation}
where
\[A=2{\vartheta}-(4{\vartheta}+1)p\mbox{ and }B =(4R-2)(4{\vartheta}+1)(1-2p).\]
Similarly,
\[f_{{\vartheta}}''(p)= 4R\,p^{4R{\vartheta}-2}(1-p)^{4R{\vartheta}+R-2} \Bigl\{C\,(1-p)^R+D\,p^R \Bigr\}\]
where
\begin{align*}
C &= 2{\vartheta}(4R{\vartheta}-1)(1-p)^2-8{\vartheta}(2{\vartheta}+1)Rp(1-p) \\
  &\quad +(2{\vartheta}+1)(4R{\vartheta}+2R-1)p^2,\\
D &= (4R-2)(4{\vartheta}+1)[(4R{\vartheta}+R-1)(1-2p)^2\\
  &\quad -2p(1-p)],
\end{align*}
and so
\begin{equation}
\label{equation2}
C\,(1-p)^R+D\,p^R=0.
\end{equation}
Combining (\ref{equation1}) and (\ref{equation2}), we get
\[(AD-BC)\,p^R=0.\]
Since $p^R\ne 0$, we have $AD-BC=0$, and so
\begin{align*}
0 &=\frac{AD-BC}{(4R-2)(4{\vartheta}+1)} \\
  & = (8R{\vartheta}+2R-1)p^2-(8R{\vartheta}+R-1)p+2R{\vartheta}.
\end{align*}
Solving this for $p$, we get two solutions
\[
p_1=\frac{8R{\vartheta}+R-1- \Delta_{R,{\vartheta}}}{2(8R{\vartheta}+2R-1)},
\]
\[
p_2=\frac{8R{\vartheta}+R-1+ \Delta_{R,{\vartheta}}}{2(8R{\vartheta}+2R-1)}.
\]
where
\[\Delta_{R,t}=\sqrt{(R-1)^2-8Rt}.\]
We see that if $8{\vartheta}R>(R-1)^2$, then the roots are not real. This reflects the fact that $f_{{\vartheta}}''(p_0)=0$ is not possible in this case.

The smaller of the two roots is the $p_1$ we are looking for, the larger $p_2$ occurs because we have neglected the two smallest
terms in (\ref{exp}) as explained above. Therefore, it is not relevant for our analysis
of $P_{{\vartheta}}(p)$.
Since $f'_{{\vartheta}}(p_1)=0$, (\ref{fdir}) implies
\begin{equation}
\label{llikk}
[2{\vartheta}-(4{\vartheta}+1)p_1](1-p_1)^R+(4R-2)(4{\vartheta}+1)(1-2p_1)p_1^R=0.
\end{equation}
Substituting the value of $p_1$ into (\ref{llikk}) and simplifying, we get (\ref{tildef}).

We cannot find a closed expression for ${\vartheta}$, but, for a given $k$, we can determine the value numerically.
We note, however, that $\lceil {\vartheta}\rceil$ (which is the quantity we want) actually is the least integer $t$ such that
\begin{equation}
\label{lik2}
\frac{2(4R-2)(4t+1)(R+\Delta )}{4tR +R-1 - (4t+1)\Delta} \Bigl(\frac{8Rt+R-1- \Delta}{8Rt+3R-1+ \Delta}\Bigr)^R \ge 1,
\end{equation}
where $\Delta=\Delta_{R,t}$ and this observation simplifies the numeric determination of $\lceil {\vartheta}\rceil$ since we do not have to
solve the equation (\ref{tildef}), but only search for $\lceil {\vartheta}\rceil$.

To prove Theorem \ref{tilde}b), we first give a couple of lemmas.
\begin{lemma}
\label{kat1} If $0<8Rt\le (R-1)^2$, then
\[ \frac{8Rt+R-1-\Delta }{8Rt+3R-1+ \Delta }  > 1-\frac{1}{2t}.\]
\end{lemma}
\begin{IEEEproof}
The function
\[ \frac{8Rt+R-1- x}{8Rt+3R-1+ x} \]
is decreasing when $x$ is increasing.
We have
\begin{align*}
\Delta^2 &=(R-1)^2-8Rt \\
&=(R-4t)^2-2R-16t^2+1< (R-4t)^2
\end{align*}
and so $\Delta <R-4t$. Hence,
\begin{align*}
&\frac{8Rt+R-1- \Delta}{8Rt+3R-1+ \Delta} > \frac{8Rt+R-1-(R- 4t)}{8Rt+3R-1+(R- 4t)}\\
&\quad = 1-\frac{1}{2t}+ \frac{4t(4t-1)+4R-1}{2t[(8R-4)t+4R-1]}>
   1-\frac{1}{2t}.
\end{align*} 
\end{IEEEproof}

\begin{lemma}
\label{kat2} If $0<8Rt\le (R-1)^2$, then
\[\frac{2(4R-2)(4t+1)(R+\Delta)}{4tR +R-1 - (4t+1)\Delta} > 8R-4.\]
\end{lemma}
\begin{IEEEproof}
We have
\begin{align}
\frac{(4t+1)(R+\Delta)}{4tR +R-1 - (4t+1)\Delta}& > \frac{(4t+1)(R+\Delta)}{4tR +R- (4t+1)\Delta} \nonumber\\
    & =\frac{R+\Delta}{R-\Delta}\ge 1. \label{ul1}
\end{align}
\end{IEEEproof}

 From Lemmas \ref{kat1} and \ref{kat2}, we see that if

\begin{equation}
\label{ll2}
8R \Bigl(1-\frac{1}{2R} \Bigr) \Bigl(1-\frac{1}{2t} \Bigr)^R\ge 1,
\end{equation}
then (\ref{lik2}) is satisfied and so $t\ge {\vartheta}$.
Taking logarithms of (\ref{ll2}), we get the equivalent expression
\[\ln (8R)+\ln\Bigl(1-\frac{1}{2R}\Bigr) + R \ln\Bigl(1-\frac{1}{2t}\Bigr)\ge 0.\]

Since $\ln (1-x)>-x/(1-x)$ for $x\in (0,1)$, we see that if
\begin{equation}
\label{ll4}
\ln (8R)-\frac{1}{2R-1} - \frac{R}{2t-1}= 0,
\end{equation}
then (\ref{ll2}) is satisfied, and so $t\ge {\vartheta}$.

Since $R=2^{k-4}$, we have $\ln (8R)=(k-1)\ln 2$. Solving (\ref{ll4}) for $t$,
we get the following relation:
\begin{equation}
{\vartheta} \le \frac{R}{2(k-1)\ln 2-2/(2R-1)}+ \frac{1}{2}. \label{tilweak2}
\end{equation}

In the proof of Lemma \ref{kat2}, we used that
$\Delta\ge 0$. However, using (\ref{tilweak2}) we get a better bound on $\Delta$ and hence a
stronger version of Lemma \ref{kat2} and a better bound on $\vartheta$.

\begin{lemma}
\label{kat3}
If
\begin{equation}
\label{nyb}
 0<t \le \frac{R}{2(k-1)\ln 2-2/(2R-1)}+ \frac{1}{2},
\end{equation}
then
\[\frac{2(4R-2)(4t+1)(R+\Delta)}{4tR +R-1 - (4t+1)\Delta} > (4R-2)(k-3).\]
\end{lemma}
\begin{IEEEproof}
By (\ref{nyb})
\begin{align}
\Delta_{R,t}^2 &= (R-1)^2-8Rt \nonumber \\
&\ge (R-1)^2-8R \frac{R}{2(k-1)\ln 2-2/(2R-1)}- \frac{8R}{2} \nonumber \\
& > R^2\Bigl(1-\frac{4}{k-1}\Bigr)^2 \label{deltaskr}
\end{align}
for $k\ge 11$.
Hence,
\[\frac{R+\Delta}{R-\Delta}\ge \frac{R+R\Bigl(1-\frac{4}{k-1}\Bigr)}{R-R\Bigl(1-\frac{4}{k-1}\Bigr)}=\frac{k-3}{2}.\]
\end{IEEEproof}

Therefore, if
\[4R \Bigl(1-\frac{1}{2R} \Bigr)(k-3) \Bigl(1-\frac{1}{2t} \Bigr)^R\ge 1,\]
then $t\ge {\vartheta}$.
Taking logarithms and solving as above, we get Theorem \ref{tilde}b) exactly as we obtained (\ref{tilweak2}) from (\ref{ll2}).

For $6\le k \le 10$, we can show that Theorem \ref{tilde}b) is true by direct computation.

To prove Theorem \ref{tilde}c), we first give another lemma.
\begin{lemma}
\label{kat1a} 
\begin{align*}
\mbox{a)}& \mbox{ For }k\ge 6\mbox{ we have } \pi<1-\frac{k-3}{2(k-1)\vartheta(k)+(k-2)}. \\
\mbox{b)}& \mbox{ For }k\ge 11\mbox{ we have } \pi^{R} < 2^{-k}.
\end{align*}
\end{lemma}
\begin{IEEEproof}
Using (\ref{deltaskr}), we get
\begin{align*}
\pi &= \frac{8R\vartheta +R-1-\Delta_{R,\vartheta} }{8R\vartheta +3R-1+ \Delta_{R,\vartheta} }\\
    &< \frac{8R\vartheta +R-\Delta_{R,\vartheta} }{8R\vartheta +3R+ \Delta_{R,\vartheta} }\\
    &< \frac{8R\vartheta +R-R[1-4/(k-1)] }{8R\vartheta +3R+ R[1-4/(k-1)] }\\
    &= 1-\frac{k-3}{2(k-1)\vartheta +(k-2)}.
\end{align*}
This is Lemma \ref{kat1a}a). Moreover, this implies that
\[\pi^R < e^{-\frac{R(k-3)}{2(k-1)\vartheta+k-2} }.\]
By Theorem \ref{tilde}b)
\begin{align*}
&\frac{R(k-3)}{2(k-1)\vartheta+k-2} \\
&\qquad > \frac{R(k-3)}{(k-1)\frac{R}{(k-2)\ln 2+\ln(k-3)-1/(2^{k-3}-1)}+2k-3} \\
&\qquad > k\ln 2
\end{align*}
for $k\ge 23$, and so $\pi^R<e^{-k\ln 2}=2^{-k}$. Direct computations show
that  $\pi^R<2^{-k}$ also for $11\le k \le 22$. Hence Lemma \ref{kat1a}b) is proved.
\end{IEEEproof}

To prove Theorem \ref{tilde}c), let $\varepsilon>0$ and $t>\varepsilon$. We get
\begin{align*}
f'&_{t-\varepsilon}(p) \\
  &= 8R(t-\varepsilon )\,p^{4R(t-\varepsilon )-1}(1-p)^{4R(t-\varepsilon )+2R} \\
  & \quad -[8R(t-\varepsilon )+4R]p^{4R(t-\varepsilon )}(1-p)^{4R(t-\varepsilon )+2R-1} \\
      &\quad +(16R-8)[4R(t-\varepsilon )+R] \\
      &\qquad \cdot [p(1-p)]^{4R(t-\varepsilon )+R-1}(1-2p) \\
      &= f'_{t}(p)-8R\varepsilon p^{4R(t-\varepsilon )-1}(1-p)^{4R(t-\varepsilon )+2R} \\
      &\quad +8R\varepsilon p^{4R(t-\varepsilon )}(1-p)^{4R(t-\varepsilon )+2R-1} \\
      &\quad -4R\varepsilon (16R-8)[p(1-p)]^{4R(t-\varepsilon )+R-1}(1-2p).
\end{align*}
Since $f'_{\vartheta}(p_1)=0$ and $(1-2p_1)/(1-p_1)=1-\pi$, we get

\begin{align*}
f'_{{\vartheta}-\varepsilon}(p_1)
      &=-8R\varepsilon (1-p_1)^{8R({\vartheta}-\varepsilon )+2R-1} \pi^{4R({\vartheta}-\varepsilon)-1 } \\
      &\qquad \cdot (1-\pi) \bigl[1+(8R-4)\pi^R \bigr].
\end{align*}

Let
\[g_t(p) = 4\,p^{4Rt+2R}(1-p)^{4Rt}   + p^{8Rt}(1-p)^{2R}. \]
Then $P_t(p)=f_t(p)+g_t(p)$.
We get
\begin{align*}
g'_{{\vartheta}-\varepsilon}(p_1) &= (1-p_1)^{8R({\vartheta}-\varepsilon )+2R-1} \pi^{4R({\vartheta}-\varepsilon)-1 } \\
      &\quad\cdot\Bigl\{[16R({\vartheta}-\varepsilon)+8R]\pi^{2R} -16R({\vartheta}-\varepsilon)\pi^{2R+1} \\
      &\quad \quad +8R({\vartheta}-\varepsilon)\pi^{4R({\vartheta}-\varepsilon )}-2R\pi^{4R({\vartheta}-\varepsilon)+1}\Bigr\} \\
      &< 8R(1-p_1)^{8R({\vartheta}-\varepsilon )+2R-1} \pi^{4R({\vartheta}-\varepsilon)-1} \\
      &\quad\cdot \Bigl\{ \bigl[ 2{\vartheta}(1-\pi)+1\bigr]\pi^{2R} +\vartheta \pi^{4R({\vartheta}-\varepsilon)}\Bigr\}.
\end{align*}

Clearly, $P'_{{\vartheta}-\varepsilon}(p_1)=f'_{{\vartheta}-\varepsilon}(p_1)+g'_{{\vartheta}-\varepsilon}(p_1)<0$
if

\begin{equation}
\label{pico1}
\bigl[ 2{\vartheta}(1-\pi)+1\bigr]\pi^{2R} \le \varepsilon(1-\pi),
\end{equation}
and
\begin{equation}
\label{pico2}
\vartheta \pi^{4R({\vartheta}-\varepsilon)} \le \varepsilon(1-\pi)(8R-4)\pi^R.
\end{equation}

Equation (\ref{pico1}) is equivalent to
\begin{equation}
\label{pico11}
\Bigl( 2{\vartheta}+\frac{1}{1-\pi}\Bigr)\pi^{2R} \le \varepsilon.
\end{equation}
We choose the $\varepsilon$ which gives equality in (\ref{pico11}), that is
\begin{equation}
\label{vardef}
\varepsilon=\Bigl( 2{\vartheta}+\frac{1}{1-\pi}\Bigr)\pi^{2R}.
\end{equation}

Equation (\ref{pico2}) is equivalent to
\begin{equation}
\label{pico12}
\frac{1}{1-\pi}\vartheta \pi^{4R({\vartheta}-\varepsilon)-3R} \le \frac{\varepsilon}{\pi^{2R}} (8R-4).
\end{equation}

By (\ref{vardef}),
\[ \frac{1}{1-\pi}< \frac{\varepsilon}{\pi^{2R}}.\]
Further, by Theorem \ref{tilde}b), 
\[\vartheta<(8R-4),\]
and, finally, $\pi<1$ and so
\[\pi^{4R({\vartheta}-\varepsilon)-3R} < 1.\]
Hence, (\ref{pico2}) is also satisfied.
Therefore 
\begin{equation}
\label{eqvar}
\theta_1\ge {\vartheta}-\varepsilon,
\end{equation}
 where $\varepsilon$ is given by (\ref{vardef}). 
By Lemma \ref{kat1a},
\begin{align}
\varepsilon &=\Bigl( 2{\vartheta}+\frac{1}{1-\pi}\Bigr)\pi^{2R} \nonumber \\
 &< \Bigl( 2{\vartheta}+\frac{2(k-1)\vartheta +k-2}{k-3}\Bigr)4^{-k}\nonumber \\
 &= \frac{k-2}{k-3}(4\vartheta+1)4^{-k}.\label{epp}
\end{align}
Combining (\ref{eqvar}) and (\ref{epp}) we get Theorem \ref{tilde}c) for $k\ge 11$. 
Direct computations show that it is true also for $6\le k \le 10$. This
 completes the proof of Theorem \ref{tilde}.

\section*{Appendix II\\ Proof of Theorem \ref{del}}

Let $\Phi_{t,k,m}$ be the number of $n$ in
\[[2^{k-1}(t+1)-2^{k-m}+1,2^{k-1}(t+1)-2^{k-m-1}]\]
 such that $S_{n,k}$ is proper. Clearly,
\begin{equation}
\label{tetsum}
\Phi_k=\sum_{t=1}^{2^{k-5}-1} \sum_{m=1}^{k-1} \Phi_{t,k,m}.
\end{equation}

\begin{lemma}
\label{tau-bounds}
Let $k\ge 6$ and $1\le m \le k-1$.

 a) If $1\le m\le \left\lfloor \frac{k-3}{2}\right\rfloor$ and $1\le t\le 2^{k-2m-3}-1$, then
\[\Phi_{t,k,m}> 2^{\frac{k+1}{2}}\sqrt{t}-1.\]

b) If $1\le m\le \left\lfloor \frac{k-3}{2}\right\rfloor$ and $t\ge 2^{k-2m-3}$, then
\[\Phi_{t,k,m}=2^{k-m-1}.\]

c) If $\left\lfloor \frac{k-3}{2}\right\rfloor+1\le m \le k-1$ and $t\ge 1$, then
\[\Phi_{t,k,m}=2^{k-m-1}.\]
\end{lemma}

\begin{IEEEproof}
By Theorem \ref{S-propek-t},
\begin{equation}
\label{tet}
\Phi_{t,k,m}\ge \lfloor \overline{\tau}_{t,k,m} \rfloor+ \lfloor \underline{\tau}_{t,k,m} \rfloor +1.
\end{equation}
In the proof of Theorem \ref{tp-kange} we showed
that if
\[\left\lfloor \frac{k-3}{2}\right\rfloor+1\le m \le k-1\mbox{ or }t\ge 2^{k-2m-3},\]
then
\[\overline{\tau}_{t,k,m}=\underline{\tau}_{t,k,m}+1=2^{k-m-2}.\]
This proves b) and c).

Now, consider $1\le m\le \left\lfloor \frac{k-3}{2}\right\rfloor$ and $1\le t\le 2^{k-2m-3}-1$.
Then $2^{k-m+2}\le 2^{k+1}$ and so
\begin{align*}
\lfloor \overline{\tau}_{t,k,m}\rfloor > \overline{\tau}_{t,k,m}-1
&=\frac{-1 +\sqrt{1+2^{k+1}(t+1)-2^{k-m+2}}}{2}\\
&\ge\frac{-1 +\sqrt{2^{k+1} t}}{2}.
\end{align*}
Similarly,
\[\lfloor \underline{\tau}_{t,k,m}\rfloor>\frac{-3 +\sqrt{2^{k+1} t}}{2}.\]
Combining these two inequalities with (\ref{tet}), a) follows.
\end{IEEEproof}

\begin{lemma}
\label{sum3}
 For $k\ge 6$ we have
\begin{align*}
\sum_{m=1}^{\left\lfloor \frac{k-3}{2} \right\rfloor}& \sum_{t=1}^{2^{k-2m-3}-1} \Phi_{t,k,m} \\
& > \frac{8}{21}\cdot 2^{2k-6}-2^{k-1}
-\frac{2^{2k-3-3\left\lfloor \frac{k-3}{2} \right\rfloor}}{21}+2^{k-1-\left\lfloor \frac{k-3}{2} \right\rfloor}\\
&\quad - \frac{2^{k-3}-2^{k-3-2\left\lfloor \frac{k-3}{2} \right\rfloor}}{3} + \left\lfloor \frac{k-3}{2} \right\rfloor.
\end{align*}
\end{lemma}

\begin{IEEEproof}
 First we see that
\begin{align*}
\sum_{t=1}^{2^{k-2m-3}-1} \sqrt{t} &= - \sqrt{2^{k-2m-3}}+ \sum_{t=1}^{2^{k-2m-3}}\sqrt{t}\\
&>- \sqrt{2^{k-2m-3}} + \int_0^{2^{k-2m-3}} \sqrt{x}\, dx\\
&= -2^{\frac{k-3}{2} -m} + \frac{2}{3} \bigl(2^{k-2m-3}\bigr)^{\frac{3}{2}}.
\end{align*}
Hence
\begin{align*}
\sum_{m=1}^{\left\lfloor \frac{k-3}{2}\right\rfloor}& \sum_{t=1}^{2^{k-2m-3}-1} 2^{\frac{k+1}{2}}\sqrt{t} \\
&>\sum_{m=1}^{\left\lfloor \frac{k-3}{2}\right\rfloor} \Bigl(\frac{2}{3}\cdot 2^{2k-4 -3m}- 2^{k-1 -m} \Bigr)\\
&= \frac{2}{21}\Bigl(2^{2k-4}-2^{2k-4-3\left\lfloor \frac{k-3}{2}\right\rfloor}\Bigr) -\Bigl(2^{k-1}-2^{k-1-\left\lfloor \frac{k-3}{2}\right\rfloor}\Bigr).
\end{align*}
Similarly,
\[\sum_{m=1}^{\left\lfloor \frac{k-3}{2}\right\rfloor}\sum_{t=1}^{2^{k-2m-3}-1} 1
= \frac{2^{k-3}-2^{k-3-2\left\lfloor \frac{k-3}{2}\right\rfloor}}{3}-\left\lfloor \frac{k-3}{2}\right\rfloor.\]
The lemma follows from these results and Lemma \ref{tau-bounds}a).
\end{IEEEproof}

\begin{lemma}
\label{sum1}
 For $k\ge 6$ we have
\begin{align*}
 \sum_{m=\left\lfloor \frac{k-3}{2}\right\rfloor+1}^{k-1}& \sum_{t=1}^{2^{k-5}-1} \Phi_{t,k,m} \\
& =2^{2k-6-\left\lfloor \frac{k-3}{2}\right\rfloor}-2^{k-5}- 2^{k-1-\left\lfloor \frac{k-3}{2}\right\rfloor}+1
\end{align*}
and
\begin{align*}
 \sum_{m=1}^{\left\lfloor \frac{k-3}{2}\right\rfloor}& \sum_{t=2^{k-2m-3}}^{2^{k-5}-1} \Phi_{t,k,m} \\
&=\frac{3}{7}\cdot 2^{2k-6} -2^{2k-6-\left\lfloor \frac{k-3}{2}\right\rfloor}+ \frac{2^{2k-4-3\left\lfloor \frac{k-3}{2}\right\rfloor}}{7}.
\end{align*}
\end{lemma}

\begin{IEEEproof}
The result follows directly from Lemmas \ref{tau-bounds}c) and \ref{tau-bounds}b) respectively.
\end{IEEEproof}

We can now combine these results into a proof of Theorem \ref{del}.

\begin{IEEEproof}
 From (\ref{tetsum}) and Lemmas \ref{tau-bounds}--\ref{sum1}, we get
\begin{align*}
\Phi_k =&\, \sum_{m=1}^{\left\lfloor \frac{k-3}{2}\right\rfloor}\sum_{t=1}^{2^{k-2m-3}-1}\Phi_{t,k,m} \\
&+ \sum_{m=\left\lfloor \frac{k-3}{2}\right\rfloor+1}^{k-1}\sum_{t=1}^{2^{k-5}-1} \Phi_{t,k,m} \\
& + \sum_{m=1}^{\left\lfloor \frac{k-3}{2}\right\rfloor} \sum_{t=2^{k-2m-3}}^{2^{k-5}-1} \Phi_{t,k,m} \\
>&\quad 2^{2k-6-\left\lfloor \frac{k-3}{2}\right\rfloor}-2^{k-5}- 2^{k-1-\left\lfloor \frac{k-3}{2}\right\rfloor}+1\\
& + \frac{3}{7}\cdot 2^{2k-6} -2^{2k-6-\left\lfloor \frac{k-3}{2}\right\rfloor}+ \frac{2^{2k-4-3\left\lfloor \frac{k-3}{2}\right\rfloor}}{7} \\ & + \frac{8}{21}\cdot 2^{2k-6}-2^{k-1}-\frac{2^{2k-3-3\left\lfloor \frac{k-3}{2}\right\rfloor}}{21}
+2^{k-1-\left\lfloor \frac{k-3}{2}\right\rfloor}\\
& -\Bigl(\frac{2^{k-3}}{3}-\frac{2^{k-3-2\left\lfloor \frac{k-3}{2}\right\rfloor}}{3}\Bigr)
+\left\lfloor \frac{k-3}{2}\right\rfloor\\
=&\, \frac{17}{21}\cdot 2^{2k-6}-\frac{55}{3}\cdot2^{k-5} \\
& + \frac{2^{2k-4-3\left\lfloor \frac{k-3}{2}\right\rfloor}}{21} + \frac{2^{k-3-2\left\lfloor \frac{k-3}{2}\right\rfloor}}{3}+\left\lfloor \frac{k-3}{2}\right\rfloor+1.
\end{align*}
Theorem \ref{del} follows from this expression.
\end{IEEEproof}


\section*{Acknowledgment
\label{sec:Ack}}

The authors are grateful to Mario Blaum for
pointing out reference \cite{DKST}. The construction of $H_k(n)$ is also essentially due to him.


\end{document}